\documentclass[prl,preprint,superscriptaddress]{revtex4-2}
\usepackage{times}
\usepackage[T1]{fontenc}
\usepackage[latin9]{inputenc}
\usepackage{color}
\usepackage{hyperref}
\usepackage{float}
\usepackage{amsmath}
\usepackage{amssymb}
\usepackage{graphicx}

\providecommand{\tabularnewline}{\\}

\begin{document}
\title{Bloch Oscillations, Landau-Zener Transition, and Topological Phase Evolution in a Pendula Array}
\author{Izhar Neder}
\email{izhar.neder@gmail.com}
\affiliation{Soreq Nuclear Research Center, Yavne 81800, Israel}
\affiliation{School of Physics and Astronomy, Tel Aviv University, Tel Aviv 69978, Israel}
\affiliation{School of Mechanical Engineering, Tel Aviv University, Tel Aviv 69978, Israel}
\author{Chaviva Sirote}
\affiliation{Department of Biomedical Engineering, Tel Aviv University, Tel Aviv 69978, Israel}
\author{Meital Geva}
\affiliation{School of Mechanical Engineering, Tel Aviv University, Tel Aviv 69978, Israel}
\author{Yoav Lahini}
\affiliation{School of Physics and Astronomy, Tel Aviv University, Tel Aviv 69978, Israel}
\affiliation{Center for Physics and Chemistry of Living Systems, Tel Aviv University, Tel Aviv 69978, Israel}
\author{Roni Ilan}
\affiliation{School of Physics and Astronomy, Tel Aviv University, Tel Aviv 69978, Israel}
\author{Yair Shokef}
\affiliation{School of Mechanical Engineering, Tel Aviv University, Tel Aviv 69978, Israel}
\affiliation{Center for Physics and Chemistry of Living Systems, Tel Aviv University, Tel Aviv 69978, Israel}
\affiliation{Center for Computational Molecular and Materials Science, Tel Aviv University, Tel Aviv 69978, Israel}
\affiliation{International Institute for Sustainability with Knotted Chiral Meta Matter, Hiroshima University, Japan}

\date{\today}

\begin{abstract}

We experimentally and theoretically study the dynamics of a one-dimensional array of pendula with a mild spatial gradient in their self-frequency and where neighboring pendula are connected with weak and alternating coupling. We map their dynamics to the topological Su-Schrieffer-Heeger (SSH) model of charged quantum particles on a lattice with alternating hopping rates in an external electric field. By directly tracking the dynamics of a wavepacket in the bulk of the lattice, we observe Bloch oscillations, Landau-Zener transitions, and coupling between the isospin (i.e. the inner wave function distribution within the unit cell) and the spatial degrees of freedom (the distribution between unit cells). We then use Bloch oscillations in the bulk to directly measure the non-trivial global topological phase winding and local geometric phase of the band. We measure an overall evolution of 3.1 $\pm$ 0.2 radians for the geometrical phase during the Bloch period, consistent with the expected Zak phase of $\pi$. Our results demonstrate the power of classical analogs of quantum models to directly observe the topological properties of the band structure, and sheds light on the similarities and the differences between quantum and classical topological effects. 

\end{abstract}

\maketitle

\newpage

\subsection*{Introduction}

Classical analogues of condensed and topological states of matter proved to be an invaluable tool in visualizing fundamental concepts and exploring the relations between quantum and classical dynamics. The topological classifications of states of systems, which originate in quantum lattice models, quantum Hall effect, quantum spin-Hall effect, topological insulators, and superconducting qubits \cite{delplace2011zak,PhysRevLett.114.114301,RevModPhys.82.3045,PhysRevLett.49.405,bernevig2006quantum,kitaev2009periodic,kane2005quantum,atala2013,Ramasesh2017,PhysRevLett.61.2015} were adapted to various classical systems such as photonic crystals~\cite{PhysRevLett.100.013904,wang2009observation,Barik2018,RevModPhys.91.015006}, phononic crystals~\cite{PhysRevLett.114.114301, PhysRevLett.103.248101,Xiao2015,Paulose2015, SirotaMSSP2021}, classical mechanical oscillators~\cite{Kane2014, Nash2015, Susstrunk2016, Huber2016, Mitchell2018, SirotaPRL2020}, and electrical resonator circuits~\cite{Hadad2018,Serra-Garcia2019}. Many of these works focused on probing stable edge channels, which provide indirect evidence for the non-triviality of the band topology.  The classification of topological invariants via direct bulk measurement, however, is limited~\cite{price2012mapping,wimmer2017experimental}, and remains a challenge in both quantum and classical systems. 

Here we report a direct measurement of a topological and geometric phase by tracking the dynamics in the bulk of a classical system: a one-dimensional array of coupled pendula. The analysis of the measurement is based on an approximate mapping between the time evolution of the classical coupled oscillators and the quantum tight-binding or discrete Schr\"{o}dinger equation of electrons on lattice potentials, as was recently demonstrated for the non-linear case~\cite{Hadad2018, DESTYL2017,MUDA2019}. The mapping is in the spirit of the mapping in Ref.~\cite{susstrunk2016classification}, but ours is strictly local. In particular, our system is mapped onto the canonical Su-Schrieffer-Heeger (SSH) model~\cite{SSH_PRL_1979, SSH_PRB_1980}, a prototypical model for topological phase transitions in one dimension. This mapping is significant, as the SSH model has two phases: one with a trivial Zak phase~\cite{berry1984,Zak1989} that is adiabatically connected to an atomic limit, and the other with a Zak phase of~$\pi$, with an obstruction to the atomic limit. In addition, a mild monotonic change in the pendula self frequency along the array is mapped into an external electric Field.

We experimentally realized such a system of $\sim50$ pendula. Using the mapping to the SSH model, we theoretically, computationally, and experimentally show that our system exhibits phenomena that are usually discussed in the context of the quantum dynamics of electrons on ultra-clean lattices. This includes Bloch oscillations~\cite{Bloch1929}, where, due to the electric field and the periodicity of the lattice, an electron's wave function oscillates in the presence of a uniform electric field instead of accelerating, and Landau-Zener (LZ) tunneling~\cite{landau1932theorie,zener1932non}, in which the electron's wavepacket can leak to a higher energy band, or be in a superposition of the two bands, due to the time-dependent energy difference between the bands. 
Finally, these observations enable us to extract the non-trivial topological phase winding of the SSH bands from bulk measurements in our system. Specifically, we use the fact that, due to Bloch oscillations, a wave-packet that is initially localized in reciprocal space samples the entire Brillouin zone during its dynamics, thereby possessing information on the geometrical phase and the Zak phase after a full Bloch period~\cite{atala2013,Rudner2017}.   

The extraction of the geometrical phase is enabled due to the classical and macroscopic nature of our system, which allows direct measurement of wavepacket dynamics and the amplitude and phase evolution of each oscillator. Specifically, we extract the geometrical phase by comparing the wavepacket phase evolution in two experiments - one with trivial and the other with topological band structure. We accurately extracted the difference of the geometrical phase evolution between these two experiments over 400 periods of the fundamental oscillation of individual pendula, which lasted about 700 $sec$, and which indeed ends in a Zak phase difference of $\pi$ after one Bloch period, as theoretically expected.

\subsection*{Classical realization of the SSH model}

\begin{figure}[b]
\begin{centering}
\includegraphics[trim=50 80 30 50, clip,width=\columnwidth]{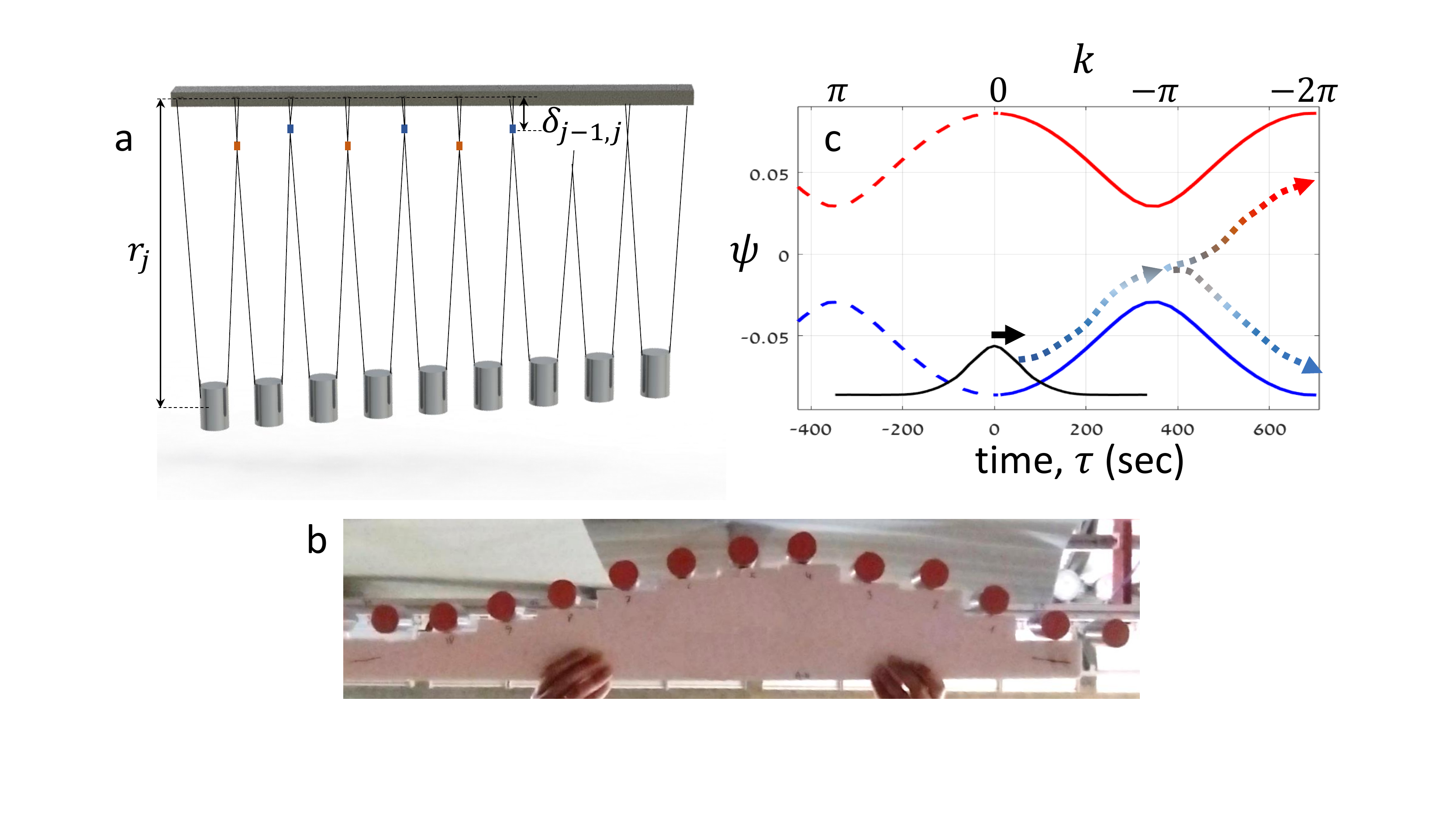}
\end{centering}
\caption{(a) Sketch of the mechanical system: pendula coupled to each other by knots (depicted by red and blue dots) connecting adjacent strings at varying and alternating heights $\delta_{j,j+1}$. The pendula lengths $r_j$ have a mild gradient according to Eq.~(\ref{eq:gradient}). (b)~The experiment started with a wave pattern implemented using a board that was cut according to the desired wave-packet. Then the board was abruptly removed and the pendula started to evolve freely.  (c) The solution to Eq.~(\ref{eq:SSH}) in Fourier space, exhibiting Bloch oscillations due to the external field: the initial wave packet (black) travels toward negative $k$ values and follows the SSH lower energy band (blue). Then at $k=-\pi$, depending on the parameters of Eq.~(\ref{eq:SSH}), the wave continues its travel either adiabatically following the lower band, by jumping through LZ diabatic transition and following the upper band (red), or by following in a superposition of the two bands. \label{fig:setup}}
\end{figure}

We constructed a one-dimensional array, schematically shown in Fig.~\ref{fig:setup}a, of $N=51$ pendula hanging using v-shaped strings from a common horizontal fixed beam. All pendula had identical mass $m=0.35\ kg$ yet varying string lengths $r_j$, ($j=1,\dots,N$), carefully-tuned to satisfy
\begin{align}
\frac{1}{r_{j}}=\frac{1}{r} \left[1+\alpha\left(j-\frac{N}{2}\right)\right]
.\label{eq:gradient}
\end{align}
The parameter $r=0.76\ m$ is the length of the central pendulum, and $\alpha=2.4\times10^{-3}$ controls the spatial gradient in string length. The accuracy in tuning the length of each pendulum was $<1\ mm$ (see Supplementary Material). 

Each pendulum is coupled to its two nearest neighbors by connecting adjacent strings by knots at alternating heights, close to the beam, as depicted in Fig.~\ref{fig:setup}a, (see also Supplementary Material). This results in weak and alternating coupling between neighboring pendula, i.e. the coupling constitutes a relatively weak perturbation to the basic pendula oscillations. The pendula oscillations are kept at small angles during the dynamics, such that we can model the system as a set of simple harmonic oscillators coupled to each other with harmonic springs of small and alternating effective stiffnesses, $\kappa_{j,j+1}$. The coupling $\kappa_{j,j+1}$ is controlled by adjusting the height of the knots that couple adjacent pendula according to
\begin{equation}
\kappa_{j,j+1}=\begin{cases}
\kappa & j\ {\rm even}\\
\kappa' & j\ {\rm odd}.
\end{cases}\label{keokoe}
\end{equation}

Details on how the knot heights determine the couplings $\kappa$ and $\kappa'$ appear in the Supplementary Material. We designed and performed three experiments, with different values of $\kappa$ and $\kappa'$. Experiments 1 and 2 were designed to observe Bloch oscillations and the change in topological phase in the adiabatic LZ regime. In Experiment 1, the couplings were $\kappa=0.07\ N/m$ and $\kappa'=0.035\ N/m$. In Experiment 2, $\kappa$ and $\kappa'$  were switched to $\kappa=0.035\ N/m$ and $\kappa'=0.07\ N/m$. Experiment 3 was designed to observe LZ tunneling and wavepackets that are in a superposition of the upper and lower bands. There, the couplings were set to $\kappa=0.082\ N/m$ and $\kappa'=0.064\ N/m$, in order to enhance the LZ transition.

The three experiments were all performed in a similar fashion. Initially, all pendula rested in their minimum energy point, except ten adjacent pendula in the bulk of the lattice, that were given an initial offset from their rest position. As shown in Fig.~\ref{fig:setup}b, these translations were given by a board, cut in advance to produce a particular initial wave pattern (see Supplementary Material).  At time $\tau=0$, the board was abruptly removed, and the system was allowed to evolve freely. The full evolution of the pendula in each experiment was video recorded from below (See Supplementary Video~\cite{supp_vid}). Finally, the position of each of the pendula as a function of time was extracted from the video.

\begin{figure}[b]
\begin{centering}
\begin{tabular}{rl}
\includegraphics[trim=10 5 35 0, clip,width=0.46\columnwidth]{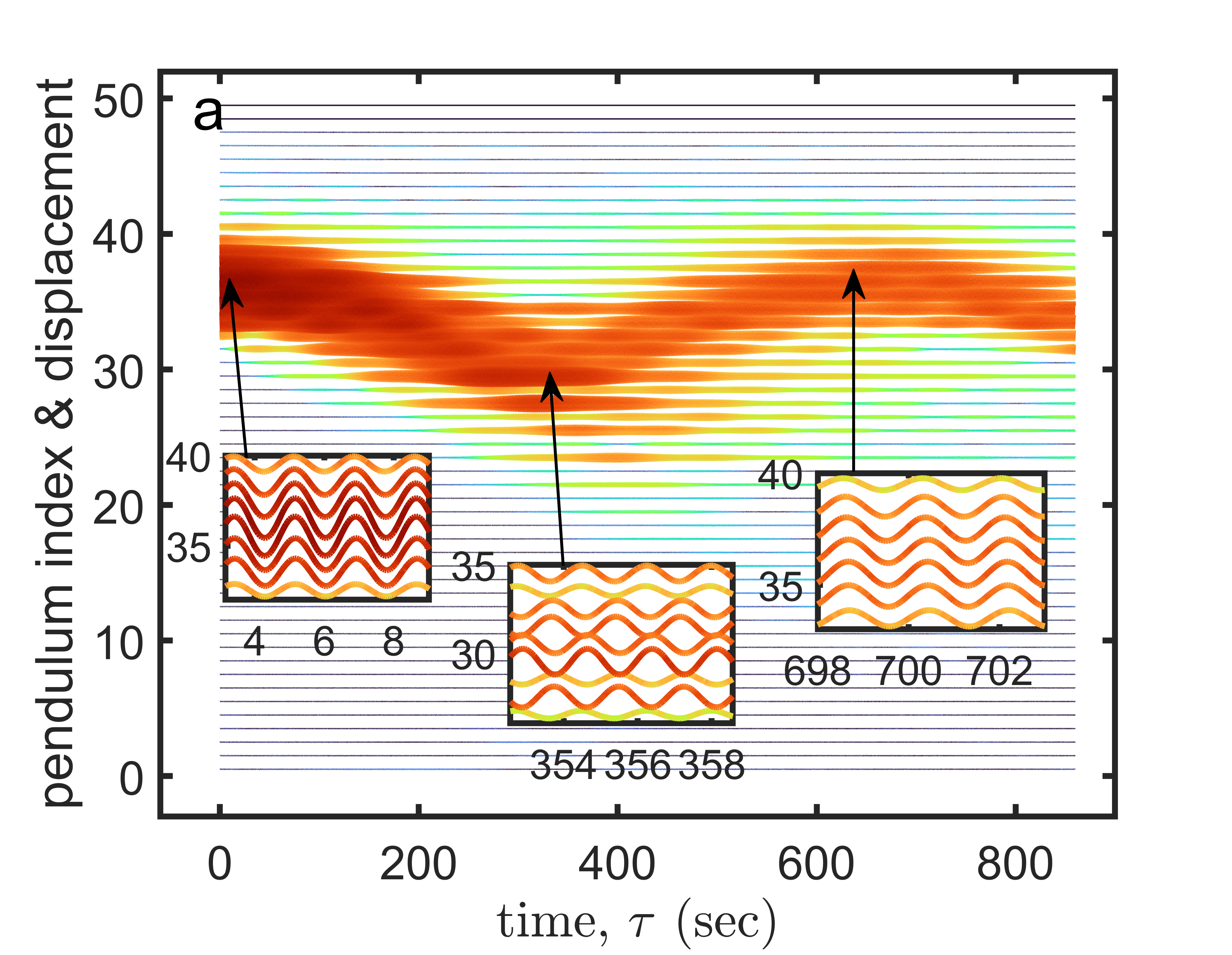} & 
\includegraphics[trim=15 5 20 10, clip,width=0.54\columnwidth]{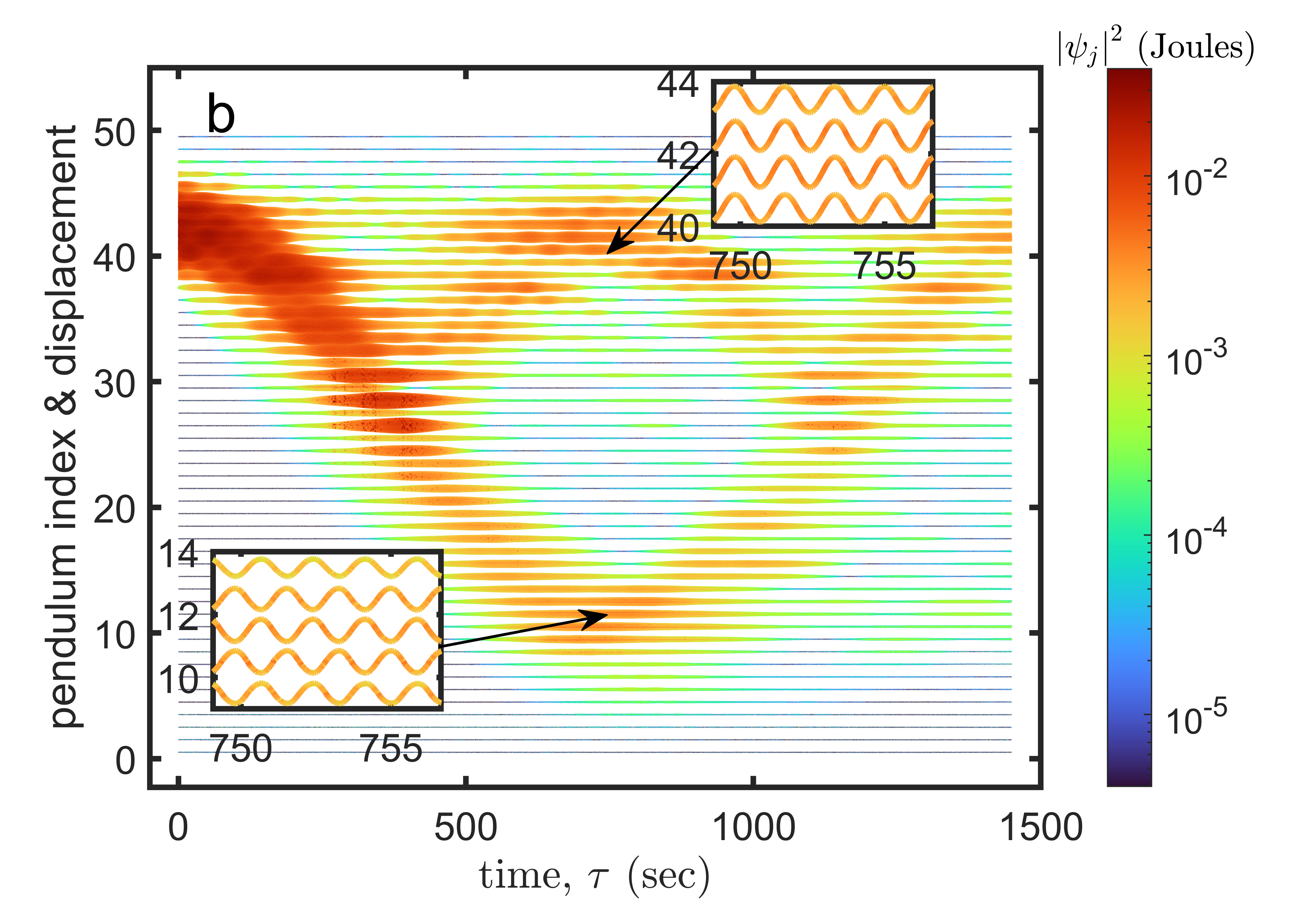}\tabularnewline
\includegraphics[trim=10 12 30 20, clip,width=0.46\columnwidth]{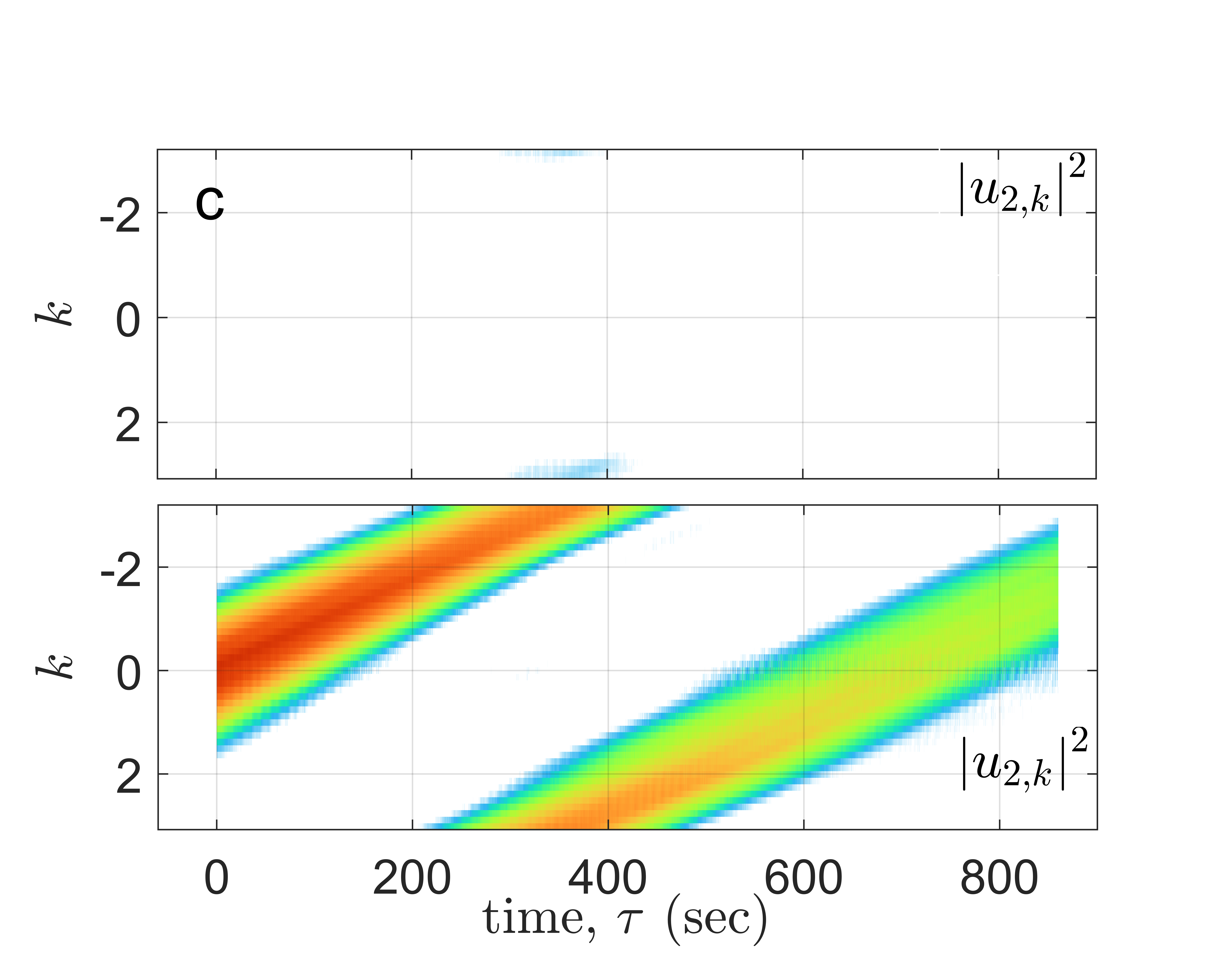} & \includegraphics[trim=20 12 20 20, clip, width=0.54\columnwidth]{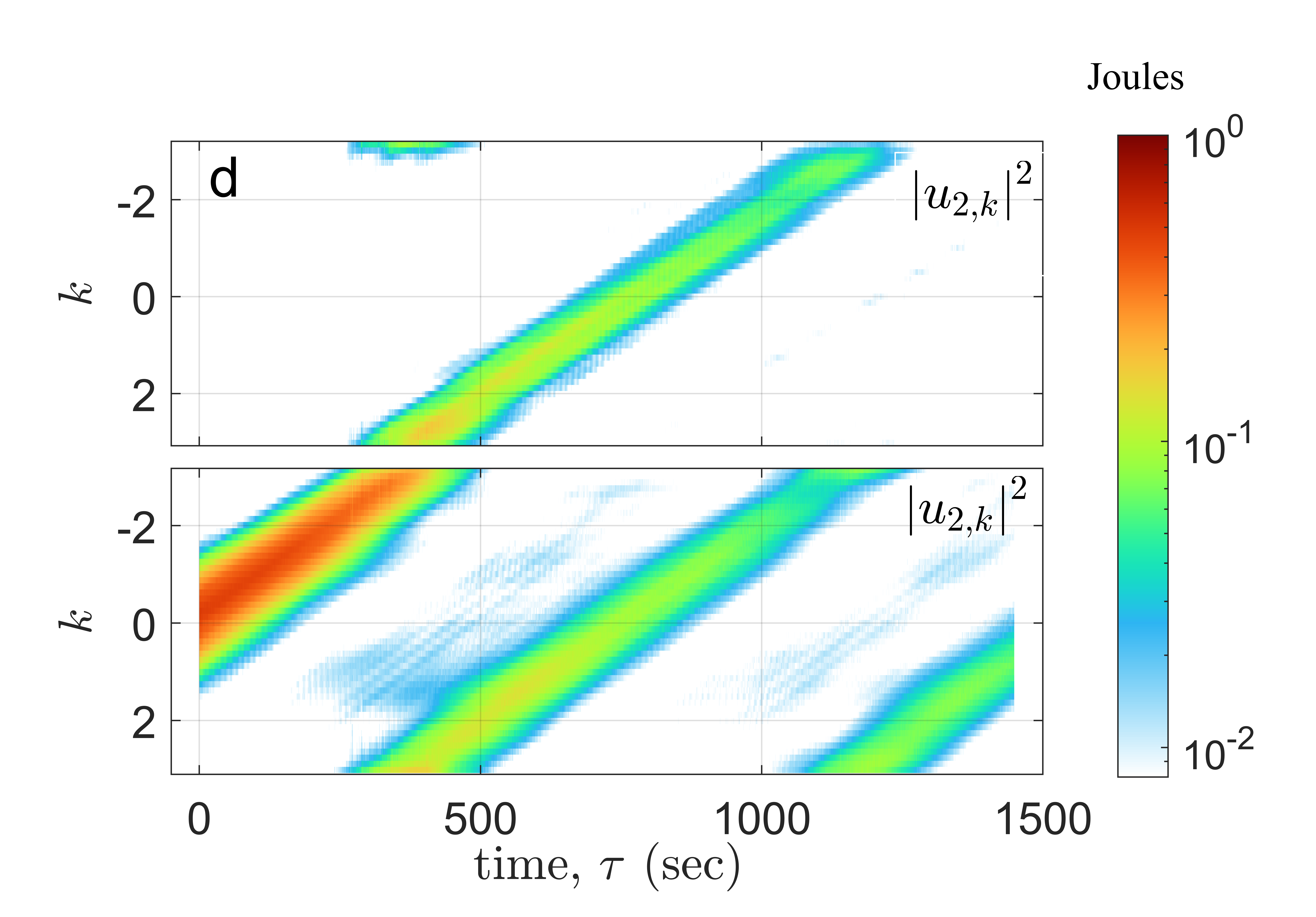}
\tabularnewline
\end{tabular}
\par\end{centering}
\caption{(a, b) The pendula displacements in Experiments 1 (a) and 3 (b). Each line shows a pendulum displacement as a function of time and is shifted vertically according to the pendulum index. Bloch oscillations are clearly seen in both experiments. In Experiment 1, the wave follows adiabatically the lower band, while in Experiment 3 after one Bloch period, the system is in a superposition of waves in the lower and upper bands. The insets show zoom-ins of the pendula displacements at certain times, in which the state of the partial wave (its mean $k$ value, and the band) can be identified (see text); (c, d)  Discrete spatial Fourier transform of the pendula motion at each time, projected onto the two theoretical eigen-states of the SSH problem, Eq.~(\ref{eq:u_12k}). These represents the partial wave in the lower (bottom) and upper (top) bands.  \label{fig:(upper-left)-tipycal}}
\end{figure}

Figure~\ref{fig:(upper-left)-tipycal} presents the results of Experiments 1 and 3, while Experiment 2 (not shown) produced results almost identical to those of Experiment 1. In Experiment 1, the energy stored in each pendulum performed slow oscillations: the wave-packet moved to lower pendula numbers and then moved back, and returned to the very same initial wavepacket at $\tau=700\ sec$, only with smaller amplitude due to $\sim$70\% energy loss due to friction. The fraction of energy that leaked to other pendula motion and did not return in the original wavepacket after $700\ sec$, was $5\%$ or less. Moreover, the inset clearly shows that in the middle of the  oscillation, at $\tau=350\ sec$ the wave was checkerboard-like: adjacent pairs of pendula were swinging out of phase, namely, the wave-packet was centered around $k=\pi$ when the unit cell is a dimer. In Experiment 3, at that middle point, the wave split into two branches, which moved toward two opposite directions at later times. However the two partial waves reached a maximal translation at $\tau=750\ sec$, and at that point a zoom-in shows that at one branch the pendula all swings in phase, while in the other branch neighboring pendula swing with alternating phases. The two branches then turned back and merged again at about $\tau=1125\ sec$.

As we now show, these observations can be understood via a mapping to the quantum SSH model, and are direct manifestations of Bloch oscillations and LZ transitions.

\subsection*{Mapping to the Quantum SSH Model}

The dynamics of the pendula system is described by $2N$ linearized Hamilton equations for the translation of each pendulum in the transverse direction, $y_{j} (\tau) = r_{j}\theta_{j} (\tau)$, where $\theta_{j}(\tau)$ is the pendulum's angle from the vertical axis, and the momentum $p_{j}(\tau)$ of each of the pendula, as a function of time $\tau$,
\begin{align}
\dot{y}_{j} & =\frac{1}{m}p_{j}\label{eq:thetadot},\\
\dot{p_{j}} & =-\frac{mg}{r_{j}}y_{j}-\kappa_{j-1,j}\left(y_{j}-y_{j-1}\right)+\kappa_{j,j+1}\left(y_{j+1}-y_{j}\right).\label{eq:pdot}
\end{align}

Experimentally, after hundreds of oscillations of the pendula, we observed significant energy dissipation of more than 50\% due to friction. However, it appeared as an overall global decay of the amplitude of all pendula and did not affect any of the effects or the measurements of the topological phases as discussed below. This can be deduced from the agreement of the experimental results with both the analytical theory and simulations. Therefore, in our theoretical analysis, energy dissipation is not considered.

Crucial to the accuracy of the mapping to the SSH model, we design the system such that the natural frequency of the pendula, $\sqrt{\frac{g}{r}}$, fulfills   $\frac{g}{r}\gg\frac{\kappa+\kappa'}{m}$, and the gradient is small, $\alpha\ll 1 $.  Namely, that $\sqrt{\frac{g}{r}}$ is significantly larger than any other frequency scale in the problem. Under these conditions, we find that if one introduces a complex dynamic variable 
\begin{align}
\psi_{j} & \equiv e^{i\omega_0\tau}u_{j} ,
\label{eq:psi_j definition}
\end{align} 
such that
\begin{align}
u_{j} & \equiv \left(\sqrt{\frac{gm}{2r}}y_{j}+i\sqrt{\frac{1}{2m}}p_{j}\right) ,
\end{align} 
and where $\omega_0\equiv\sqrt{\frac{g}{r}}+\sqrt{\frac{r}{g}}\frac{1}{m}\left(\kappa+\kappa'\right)$, then to a very good approximation, $\psi_j$ evolves according to (see derivation in Supplementary Material):
\begin{align}
i\dot{\psi_{j}} & \approxeq\frac{Ea}{2}\left(j-\frac{N}{2}\right)\psi_{j}-t_{j-1,j}\psi_{j-1}-t_{j,j+1}\psi_{j+1}\label{eq:SSH} .
\end{align}
This equation is identical to the Schr\"{o}dinger equation for the SSH model with lattice constant $a$ in the presence of an electric field $E$. The  alternating hopping terms are related to the system's parameters by
\begin{align}
t_{j,j+1} & =\begin{cases}
\sqrt{\frac{r}{g}}\frac{1}{2m}\kappa \equiv t & j\ {\rm even},\\
\sqrt{\frac{r}{g}}\frac{1}{2m}\kappa' \equiv t' & j\ {\rm odd}.
\end{cases}\nonumber\\
\label{eq:Mapping-ssh}
\end{align} 
Given Eq.~(\ref{eq:SSH}), the combination 
\begin{align}
Ea & = \sqrt{\frac{g}{r}}\alpha \label{eq:Mapping-ssh2}
\end{align}
is related to the Bloch frequency by $Ea=\frac{dk}{d\tau}$, where $-\pi<k<\pi$ is the dimensionless wave vector that corresponds to the discrete Fourier transform with respect to unit cell numbers (note that the unit cell is a dimer). The accuracy of Eq.~(\ref{eq:SSH}) depends on the original system's parameters; Specifically, Eqs.~(\ref{eq:thetadot}) and (\ref{eq:pdot}) have two small parameters. The first is $\alpha$, and the second can be defined as $\epsilon\equiv\sqrt{\frac{g\left(\kappa+\kappa'\right)}{rm}}$, which is the ratio between the pendulum and the couplings self frequencies. Both should be made as small as possible for a faithful mapping to the SSH model.

\subsection*{Experimental Measurement of Bloch Oscillations and LZ Transition}

In order to compare the phenomena seen in Fig.~\ref{fig:(upper-left)-tipycal},a,b to the prediction of Eq.~(\ref{eq:SSH}) we analyse them in Fourier space by performing two Fourier transforms over the odd-$j$ ``a'' sites and the even-$j$ ``b'' sites. On the theoretical level,  Eq.~(\ref{eq:SSH}) with zero electric field leads to two energy bands with eigenvalues $\omega_k=\pm v_k$  and eigenstates in the $(a,b)$ dimer inner space   $\xi_{k,1/2}\equiv\frac{1}{\sqrt{2}}(1,\pm e^{i\varphi_k})^{T}$ where $v_ke^{i\varphi}=t+t'e^{-k}$ (see Supplementary Material for details). Due to the small electric field, during the Bloch oscillation, an initial wave-packet changes the value of its central wavenumber $k$ in momentum space at a constant rate, periodically sampling the whole Brillouin zone $-\pi<k<\pi$. At $k=\pi$ the gap is minimal, and depending on the rate $\frac{dk}{d\tau}=Ea$, the wave packet evolution can vary, from following adiabatically the lower band, to jumping to the upper band through the LZ transition, to being in a superposition of the two bands; see illustration in Fig.~\ref{fig:setup}c. 

The Bloch oscillations observation is more convincing in $k$-space with respect to the unit cell (the dimers) number. Thus, each of the three experiments was further analyzed by taking the complex values of $u_j$ obtained from Eq.~(\ref{eq:psi_j definition}) using the measured values of $y_j$ and $p_j$,  and performing, at each point in  time, two discrete Fourier transforms of the wavefunctions in the odd ``a'' sites $u_{2l+1}$ and of the even ``b'' sites $u_{2l+2}$ and defining 
\begin{align}
u_{k,a} \equiv\sum_{l=0}^{N/2-1}{u_{2l+1}e^{-ilk}},\quad
u_{k,b} \equiv\sum_{l=0}^{N/2-1}{u_{2l+2}e^{-ilk}}.\label{eq:u_abk}
\end{align}
We project the resulting vector $(u_{k,a},u_{k,b})^{T}$ onto the analytically derived lower- and upper-energy eigenstates of the SSH model $\xi_{k,1/2}$
\begin{align}
u_{k,1} & \equiv \frac{1}{\sqrt{2}}\left(u_{k,a}+e^{-i\varphi_k}u_{k,b}\right),\nonumber\\
u_{k,2} & \equiv  \frac{1}{\sqrt{2}}\left(u_{k,a}-e^{-i\varphi_k}u_{k,b}\right),\label{eq:u_12k}
\end{align}
where $\varphi_k$ is estimated from the experimental $\kappa$ and $\kappa'$, and their mapping to $t$ and $t'$.  The result is shown in Fig.~\ref{fig:(upper-left)-tipycal}c for Experiment 1 and in Fig.~\ref{fig:(upper-left)-tipycal}d for Experiment 3. One can see that the wave-packet started its evolution almost entirely in the lower band.  Indeed, the initial condition $u_{j}(\tau=0)$ implemented by the cut board was designed to be a Gaussian in the lower band, centered at $k=0$ with $\Delta k\approx0.35$ small compared to the Brillouin zone dimensionless size of $2\pi$. 

At early times, the Gaussian in $k$-space stayed in the lower band and shifted to lower values of $k$ at a constant rate, until reaching the middle of the Bloch cycle, corresponds to $k=\pm\pi$, as can be clearly observed from the zoom-in in the insets of the corresponding figures in $j$-space. The wave-packet continued to evolve along the lower band in Experiment 1 and 2, whereas in Experiment 3 the wave split and some of its amplitude was transferred into the upper-band via a LZ transition. This could be easily spotted in $j$-space, because the continuing Bloch oscillation at later times made the two parts of the wave-packet move in different directions, due to the opposite signs of the group velocities $\frac{d\omega_k}{dk}$ of the two bands. Furthermore, the zoom-ins in $j$-space clearly show the two different inner dimer states of the two bands at $k=0$ -- the state (1,-1) for the upper band and the state (1,1) for the lower band. In this respect, the real space evolution of the wave-packets was coupled to the pseudo-spin degree of freedom defined by the sublattice (i.e, the upper and lower bands) -- similar to the situation in the Stern-Gerlach experiment. 

\subsection*{Validation of the Mapping with Numerical Simulations}

Simulating the system can further demonstrate the strength and robustness of the mapping for displaying quantum-analog effects in classical mechanical systems, over a wide range of parameter values.  We simulated the system of coupled pendula by solving numerically Eqs.~(\ref{eq:thetadot},\ref{eq:pdot}) for various sets of parameters $N$, $\alpha$, $r$, $\kappa$, and $\kappa'$. We used the 4-5 Runge-Kutta method with both absolute and relative tolerances of $10^{-9}$.

\begin{figure}[t]
\begin{centering}
\includegraphics[trim=5 3 33 11, clip,width=0.49\columnwidth]{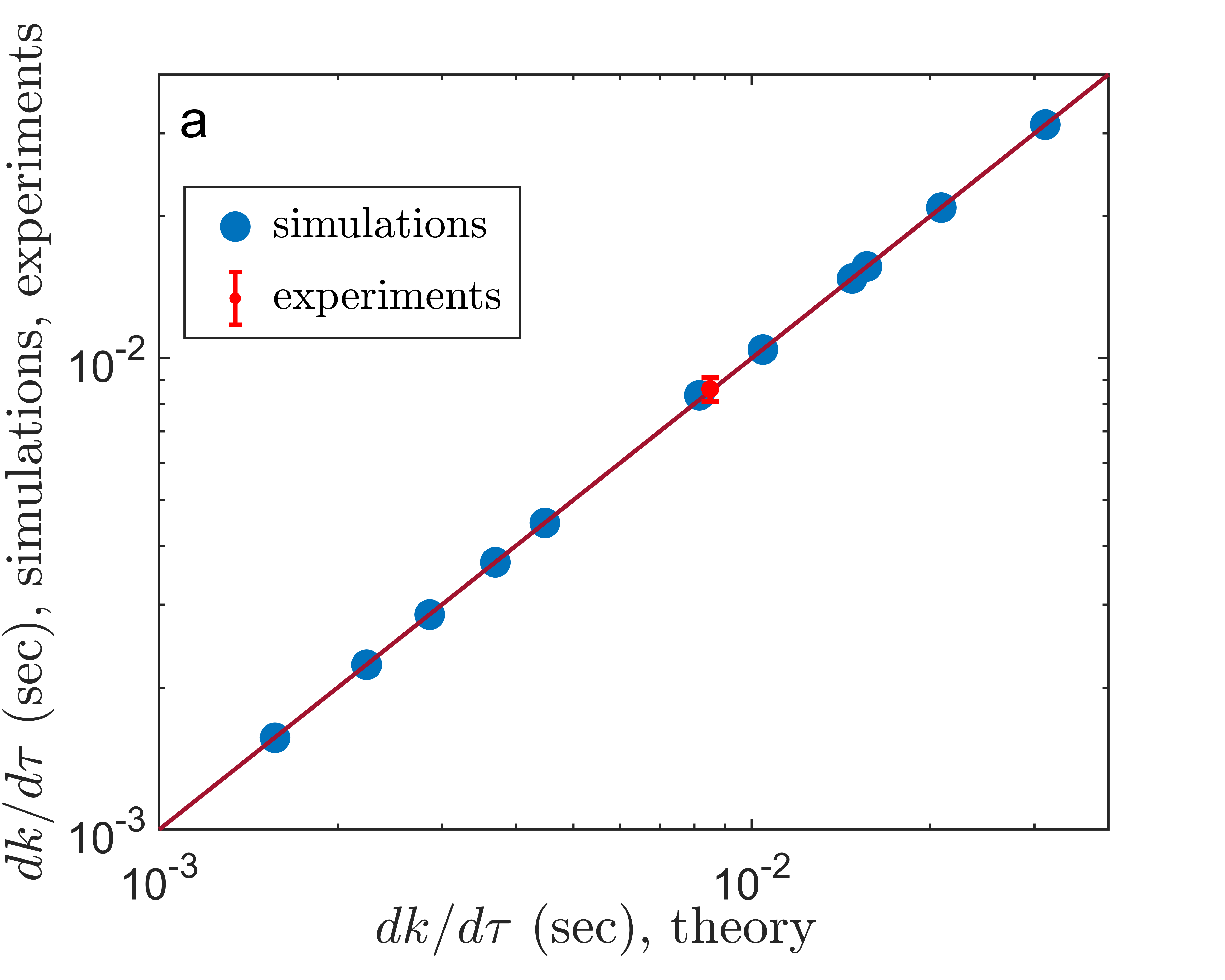}
\includegraphics[trim=5 3 33 11, clip,width=0.49\columnwidth]{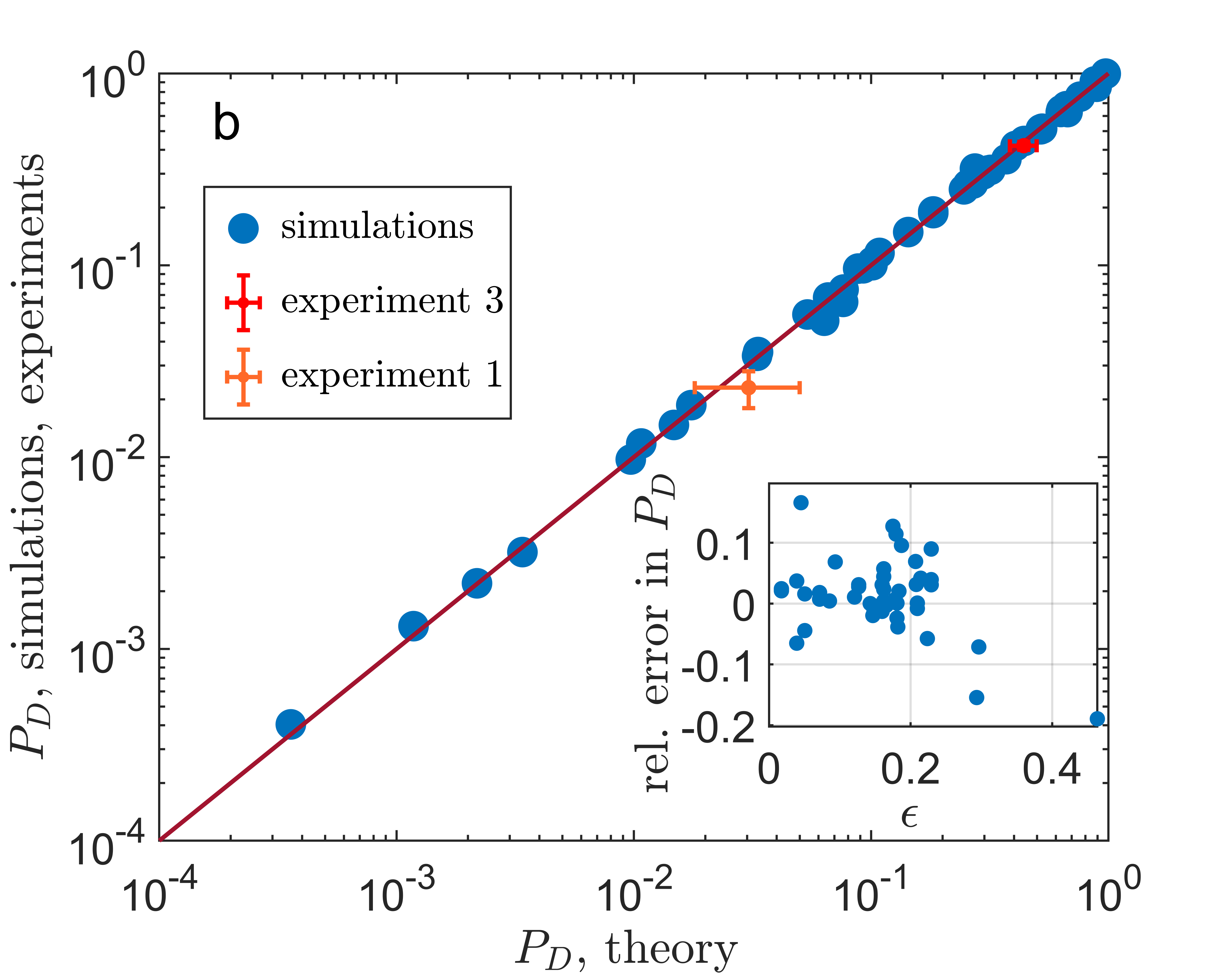}
\par\end{centering}
\caption{(a) The simulated rate of Bloch oscillations compared with the theoretical prediction $\frac{dk}{d\tau}=\sqrt{\frac{g}{r}}\alpha$ for a significant range of the values of the parameters $g$, $r$, and $\alpha$. The solid line marks the identity. The experimentally measured value (red) of the Bloch period agrees with the theoretical prediction, with the error-bar marking the spread between Experiments 1, 2, and 3. (b) The LZ diabatic transition probability as extracted from experiments and simulations, Eq.~(\ref{eq:Pd}), vs. the theoretical prediction, Eq.~(\ref{eq:LZ-pendulum}). The simulations covered variations in all parameters, $g$, $r$, $\kappa$, and $\kappa'$. The solid line marks the identity. Inset: relative deviation of the simulated values from the prediction. For reasonable values of $\epsilon=\sqrt{r(\kappa+\kappa')/gm}$, which control the accuracy of the mapping of the pendula system to the SSH model, the deviation is 5\% or less.\label{fig:(left)-The-Bloch}}
\end{figure}

We checked the accuracy of the mapping by quantitatively predicting the details of Bloch oscillations and the LZ transition. Figure~\ref{fig:(left)-The-Bloch}a verifies the validity of the theoretical expression for the Bloch frequency $\frac{dk}{d\tau}=\sqrt{\frac{g}{r}}\alpha$, by showing excellent agreement as the parameters $\alpha,g,r$ were varied over considerable ranges. Other features of the oscillations, such as their amplitude, are also well predicted by the mapping to the SSH model, see Supplementary Material for details.  

The LZ diabatic transition probability is given by $P_{D} = \exp\left(-2\pi V^{2}/d\right)$, where $V$ is the off-diagonal element and $d$ is the rate of change of the diagonal part of the Hamiltonian in the LZ model~\cite{landau1932theorie,zener1932non}. Given the energy bands in the SSH model and the mapping in Eq.~(\ref{eq:Mapping-ssh}), we can express $P_D$ using the pendula parameters (see derivation in the Supplementary Material), 
\begin{align}
P_{D} = \exp\left[-\frac{\pi r\left(\kappa-\kappa'\right)^{2}}{2mg\alpha\sqrt{\kappa\kappa'}}\right] . \label{eq:LZ-pendulum}
\end{align}
Figure~\ref{fig:(left)-The-Bloch} shows results of simulations of Eqs.~(\ref{eq:thetadot}) and (\ref{eq:pdot}) using $N=200$ pendula, using an initial Gaussian wavepacket in the lower band, for various values of $g,r,\kappa$ and $\kappa'$. The relative energy transfer to the upper band after one Bloch oscillation, at $\tau=T_B=2\pi/\frac{dk}{d\tau}$, observed in the simulations,
\begin{equation}
P_{D,\rm{sim}}\approx \frac{\sum_{k}\left|u_{k,2}(\tau=T_B)\right|^{2}}{\sum_{k}\left|u_{k,1}(\tau=0)\right|^{2}}
\label{eq:Pd}   
\end{equation}
was compared to the theoretical expression in Eq.~(\ref{eq:LZ-pendulum}). The agreement between the simulation and the analytical formula extends over almost three orders of magnitude with no fitting parameters and over a considerable range in $g,r,\kappa,\kappa'$ and $\alpha$ thus demonstrating the robustness and insensitivity of the mapping to changing parameters. When $\epsilon$ is less than 0.2, the error in predicting the LZ transition probability is a few percent or less. As a side note, this accuracy is why both the analytical calculations and the simulations were proven to be useful in designing the experiments in this work. In Fig.~\ref{fig:(left)-The-Bloch} we also present the measured values of $Ea$ and of $P_D$ from our experiments, again with good agreement to the theoretical results. Error-bars were estimated from the spread of the results in the three experiments and from the accuracy of determining the parameters in the experiment, see the Supplementary Material for more details.

\subsection*{Geometrical Phase Evolution and Band Topology}

When the values of $\kappa$ and $\kappa'$ are switched (such was the situation between Experiments 1 and 2), the topology of the bands of the SSH Hamiltonian switches from trivial to non-trivial. The topology of the band is imprinted in the evolution of the wave-packet in the form of a global $\pi$ phase difference after one Bloch period. Contrary to the situation in the quantum case, the global phase of the classical pendula oscillations variables $u_{j}$ (and of $\psi_{j}$) was easily measurable. Therefore, in our classical system, it is possible to fix a gauge and extract the whole evolution of the geometrical phase as a function of the quasi-momentum $k$ during the Bloch oscillation period that leads to this $\pi$ phase shift. This is achieved by extracting the phase of the complex wave function component in Fourier space, $u_{k(t),1}$ in Fig.~\ref{fig:(upper-left)-tipycal}c, at the maximum of the wavepacket  as a function of $\tau$.

To understand the relation between this phase of the wavepacket maximium and the topological phase, we solve the dynamic equation of the SSH model in the limit of adiabatic evolution. In the Supplementary Material, we show that in the case of an adiabatic evolution in the lower band, the solution of Eq.~(\ref{eq:SSH}) is such that the maximum of the wave packet moves in momentum space along the line $k(\tau)=(-Ea\tau +\pi)\ mod\  2\pi -\pi$ (the wavepacket maximum in Fig.~\ref{fig:(upper-left)-tipycal}c).  Along this line, $u_{k(\tau),1}$ has an additional phase term $e^{i\phi(k,\tau)}$, which is added to the basic oscillations $e^{i\omega_0t}$ and is given by 
\begin{equation}
\left.\phi\right|_{k=-Ea\tau}=\phi_{0}-\frac{N}{4}Ea\tau-\int_{0}^{\tau}\omega_{k'=-Ea\tau'}d\tau'+\frac{1}{2}\left.\varphi_{k}\right|_{0}^{-Ea\tau}.\label{eq:phase evolution}
\end{equation}
Only the last term, which is the geometrical phase evolution from wave number $k=0$ to $k=-Ea\tau$, changes between Experiments 1 and 2, namely when switching $\kappa\leftrightarrow\kappa'$. This is because in the first term, $\phi_{0}$ depends only on the initial condition,  the second term is related to the global potential being centered around the $l=N/4$ unit cell, and the third term is the dynamic phase evolution along the line $k'=-Ea\tau'$, $\tau'\in[0,\tau]$. This expression depends on the specific choice of gauge -- the choice of the global phase for the eigenstate $\xi_{k,1}\equiv\frac{1}{\sqrt{2}}(1,e^{i\varphi_{k}})^{T}$ at every $k$, as manifested in Eq.~(\ref{eq:u_12k}) above. We conclude that the only difference in  the  phase $\phi$ between Experiments 1 and 2 originates from the changes in the geometrical phase.

\begin{figure}[t]
\begin{centering}
\includegraphics[trim=0 40 0 90, clip,width=0.7\columnwidth]{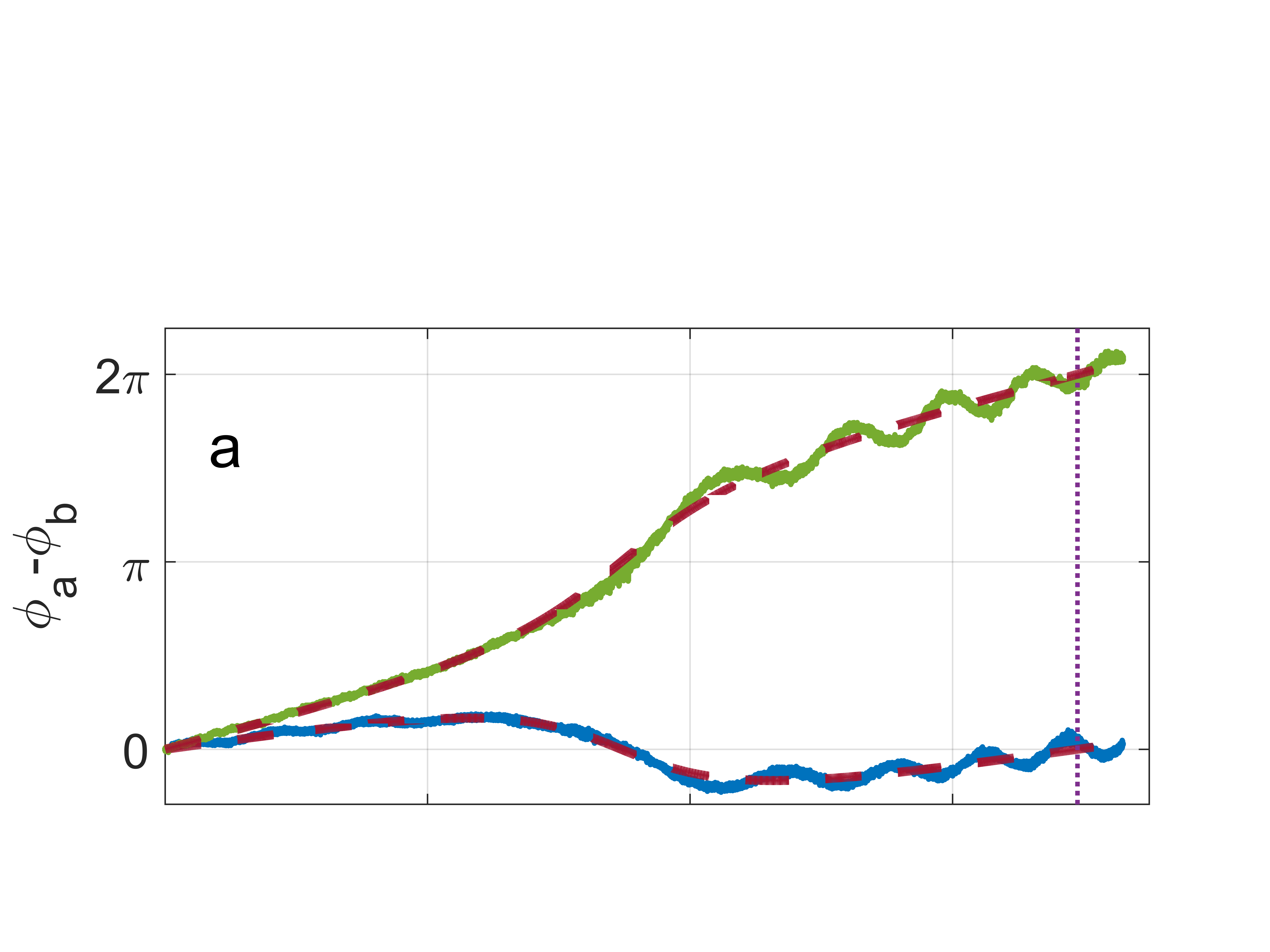}
\includegraphics[trim=0 0 0 100, clip,width=0.7\columnwidth]{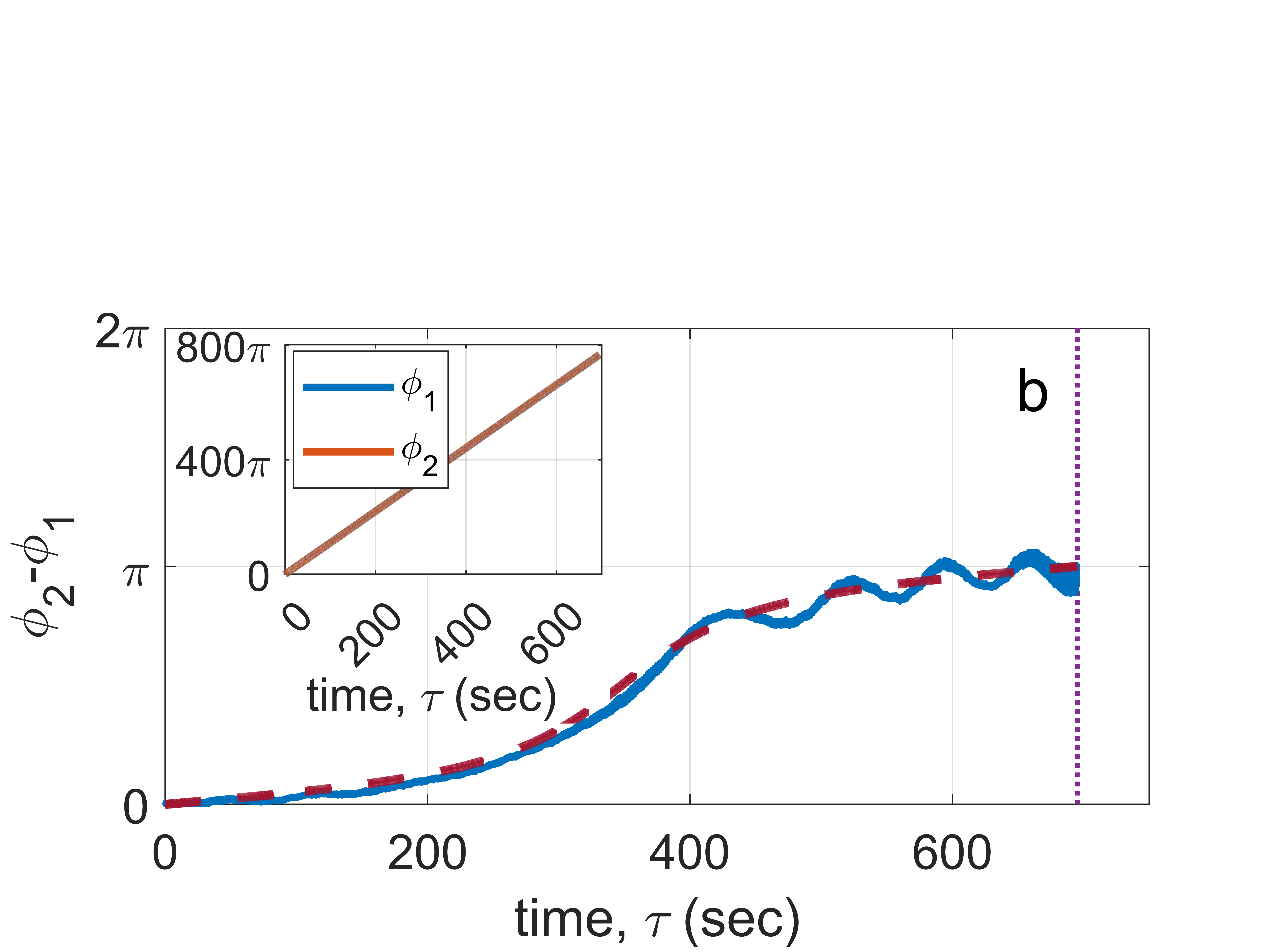}
\par\end{centering}
\caption{(a) Phase difference between the dimer components, $\arg\left(u_{b,k(\tau)}\right)-\arg\left(u_{a,k(\tau)}\right)$ at $k(\tau)=-Ea\tau$, as extracted from the pendula displacements in Experiments 1 (dark green line) and 2 (blue line). The dashed red lines show the theoretical curves $\arg(\kappa+\kappa'e^{iEa\tau})$, where in Experiment 1, $ \kappa=0.0035\, N/m,\kappa'=0.07\, N/m$ and in Experiment 2, $\kappa=0.07\, N/m,\kappa'=0.035\, N/m$. The accuracy of the couplings is about $5\%$ (1 $mm$ in the knots' heights). The purple dotted line marks the Bloch oscillation period $T_{B}$. The winding number of the phase after a full Bloch oscillation clearly shows that in Experiment 1 the set-up was in the non-trivial topological phase, while in Experiment 2 it was in the trivial topological phase. (b) Difference in global phase evolution of $u_{a,-eE\tau}$ between Experiments 2 and 1, where the full phase evolution from the two experiments are shown in the inset. After 700 $sec$, and about 2500 radians of phase evolution of the pendula swings, the difference between the phases reached $\pi$ to very good accuracy. The dashed red line is the theoretical prediction $\frac{1}{2}\arg(\kappa'+\kappa e^{-iEa\tau})-\frac{1}{2}\arg(\kappa+\kappa'e^{-iEa\tau})$, with $\kappa=0.035\, N/m,\kappa'=0.07\, N/m$.\label{fig:(left)Geometrical phase evollutions}}
\end{figure}

Figure~\ref{fig:(left)Geometrical phase evollutions}b shows the phase difference of $u_{k(\tau),1}\left(\tau\right)$ along the line $k(\tau)=-Ea\tau$ (the maximum of the wave packet) in the two experiments. The complex values of $u_{a,-eE\tau}$ and $u_{b,-eE\tau}$ in each experiment were extracted from the Fourier transform of the measured pendula displacements, followed by 2D interpolation in Fourier space. While the phase in each experiment alone completed about 2500 radians during the first Bloch period (inset), the evolution of the phase \textit{difference} between the two experiments followed to a very good approximation the theoretically predicted difference of the geometrical phase evolution, $\frac{1}{2}\arg(\kappa+\kappa'e^{-iEat})$, and completed a $\pi$ phase shift when completing a full Bloch oscillation, within a measurement accuracy of 0.2 radians. Note that the phase difference is gauge independent, as long as the same gauge is used in both experiments. 

Information about the local geometry and global topology of the band can also be extracted from a single experiment by comparing the phase evolution of $u_{k,a}\left(\tau\right)$ and $u_{k,b}\left(\tau\right)$. The phase difference of the two entries of $\xi_{k,1}$ equals $\varphi_{k}$, independent of the choice of gauge. In Fig.~\ref{fig:(left)Geometrical phase evollutions}a we plot the phase difference between $u_{k,a}\left(\tau\right)$ and $u_{k,b}\left(\tau\right)$ in the two experiments. This phase difference is shown to follow the theoretical curve $\arg(\kappa+\kappa'e^{-iEat})$, which results in a different winding number for switching $\kappa$ and $\kappa'$. The small ripples of the experimental curves, which appear to dominate the accuracy in measuring the geometrical phase and the winding are attributed to a small but finite LZ diabetic transition (1-2$\%$). Similar ripples were seen in simulations that we performed with similar conditions and can, in principle, be improved by future experiment with e.g. a larger band gap.

\subsection*{Discussion}

We have shown that a system of coupled pendula with alternating couplings and a mild gradient in the pendula's lengths obeys remarkably accurately the Schr\"{o}dinger equation of the topological SSH model in an external electric field. As such, its dynamics exhibits Bloch oscillations, LZ transitions that are followed by entanglement between the band and spatial degrees of freedom, and the appearance of the non-trivial topological phase of the bands. These features that are usually attributed to microscopic quantum systems now appear on a macroscopic classical system, which is easily observed in full detail through direct measurement of the evolution of its ``wave function''. This enabled us to quantify the Bloch oscillations rate and the LZ diabatic transition probability that both perfectly agree with theoretical predictions. Finally, to our knowledge, we are the first to directly extract the topology of the bands from an experimental measurement in the bulk of a macroscopic system -- both within one evolution of the system by measuring the relative phase of the component of the dimer, and by comparing the phase evolution in two experiments performed on the topological and the trivial states of the system. Our measurement of the Zak phase was remarkably accurate -- an accuracy of $0.2$ radians after $800\pi$ radians evolution -- a relative error of less than $10^{-4}$. This enabled us to easily observe the accumulated Zak phase of $\pi$.

The fact that these phenomena could be accurately measured in a simple, macroscopic, mechanical system is not trivial in several aspects; First, the mapping we used is local but approximate, in contrast to the one used in Ref.~\cite{susstrunk2016classification} for the classification of the topological phases, which was exact but not local. Here we maintained the locality in order to have a simple observation of the wave-function evolution. Second, we proved that, unlike the quantum case, the significant dissipation of energy to heat in our experiment did not affect the coherence of the wave. Moreover, we note that there were additional degrees of freedom in the motion of each pendulum -- such as motion along the axis of beam (on top of the oscillations in the perpendicular direction that we focused on), and rotation of each pendulum due to the unbalanced force moment acting on each pendulum from its coupled neighbours. Apparently those did not affect the coherence either, even to the above mentioned degree of accuracy.

These results open the way to quantitatively explore further quantum-like phenomena in classical systems. Future theoretical and experimental works can explore the non-linear nature of physical pendula. Another possible direction of study is to examine the effect of point defects or imperfect lattices on the topological effects. In addition, note that the SSH one-dimensional topological insulator is only one example among many quantum lattice models with non-trivial features; our work can be generalized quite naturally to other systems by considering different higher-dimensional lattices of oscillators. This includes single and double-layer honeycomb lattices, and 2D or even 3D topological insulators. 

\subsection*{Acknowledgments}

We thank Ari Sirote, Baruch Meirovich, Dafna Shokef, Hadas Shokef, Maital Silver, Raziel Katz, Tomer Sigalov and Yaara Shokef for technical assistance, and Lea Beilkin, Eran Sela, Nissim Ofek, and Yakir Hadad for useful discussions. This research was supported in part by the Israeli Ministry of Science and Technology Grant No. 3-15671. RI is supported by the Israeli Science Foundation under grant No. 1790/18 and U.S.-Israel Binational Science Foundation (BSF) Grant No. 2018226. YS is supported by the Israeli Science Foundation under grant No. 1899/20.

\bibliographystyle{unsrt}

\bibliography{refs}

\clearpage 

\section*{Supplementary Material}

\subsection*{Experimental design} 

\begin{figure}
\begin{centering}
\includegraphics[trim=40 350 10 100,clip,width=\columnwidth]{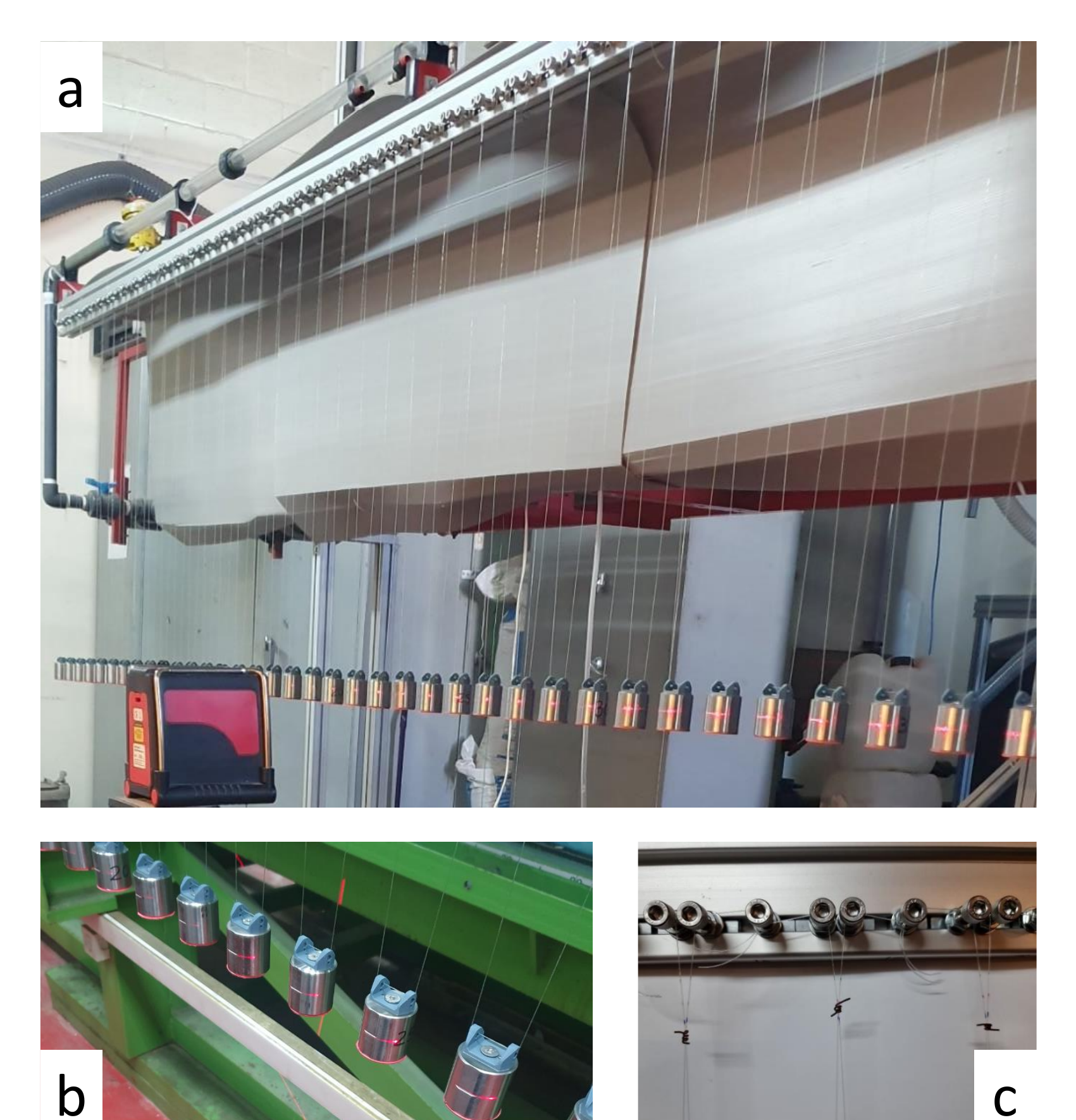}  \par\end{centering}
\caption{(a)~The lengths calibration process was done using a mildly tilted laser level. (b)~Picture of the use of the groves on the pendula weights in the leveling process with the laser level. (c)~A close-up of the screws on the beam and the coupling knots. \label{fig:system_setup}}
\end{figure}

The experimental system, shown in Fig.~\ref{fig:system_setup} consisted of a beam to which each pendulum's string (a $0.5\,mm$ fishing nylon wire) was attached by two positioning screws and one rotating screw for fine adjustment of the string's length. The nylon wire was wrapped several times on the rotating screw and was then fine-tuned by rotating the screw, after which a nut was fastened to hold the screw from rotating. The knots used for the couplings were small metal wires that were wrapped around the two adjacent pendula's nylon strings 3-4 times and held them together tight enough during the experiment, but still allowed us easy calibration by sliding them (against the static friction) up and down at will. For the calibration of the lengths of the strings and heights of the knots we used a laser level and a laser distance meter, both with an accuracy of $1\,mm$, in the following method: We first ran simulations from which we prepared a table of the desired lengths $r_j$ and couplings $\kappa\ ,\kappa'$, which led to a list of knots heights $\delta_j$  through Eq.~(\ref{eq:kappa_delta}). We calibrated the pendula in groups of 15-20 pendula at a time, where for each group we measured 3-4 pendula with the distance meter (from the beam to the weight), and for the rest we used the laser level, which was slightly tilted to go through a $1\ mm$ groove that was made in the middle of the weight of each pendulum (including the 3-4 pendula that were measured previously with the laser meter), see Fig.~\ref{fig:system_setup}. This calibration method gave us about $1\ mm$ accuracy in the pendula's absolute length's but also significantly reduced the error in the difference in lengths in adjacent pendula due to the common use of the level. This was important for a clear Bloch oscillations observation (see below elaboration on the effect of fluctuations in the pendula lengths). The pendula were given initial displacements by a cut board, shown in Fig.~\ref{fig:setup}b, and were video recorded from below (see~\cite{supp_vid}) by several smartphones. The video files from the different smartphones were synchronized by an audio signal. The position of each pendulum as a function of time was extracted from the images by identifying the circular shapes of the pendula, when viewed from below. Pendula displacements were calibrated by scaling the images according to the known weight diameter.

\subsection*{Mapping the classical pendula dynamics to the SSH model}

We start from the classical equation of motion, Eq.~(\ref{eq:pdot}), and introduce the complex dynamical variable, 
\begin{align}
u_{j} & \equiv\sqrt{\frac{gm}{2r}}y_{j}+i\sqrt{\frac{1}{2m}}p_{j}\label{eq:u_j definition} .
\end{align}
Note that  $\left|Re(u_j)\right|=\sqrt{E_{p,j}}$ and $\left|Im(u_j)\right|=\sqrt{E_{k,j}}$ are roughly related to the potential and kinetic energies, respectively, stored in pendulum $j$, up to the small changes $r-r_j$, and the energy of the weak couplings between neighboring pendula. A wave-packet evolution conserves energy, which mimics the unitary evolution of a wave function in quantum mechanics, that conserves total probability. To progress towards finding a Schr\"{o}dinger-like equation for $u_{j}$, we multiply Eq.~(\ref{eq:thetadot}) by $i\sqrt{\frac{gm}{2r}}$ and Eq.~(\ref{eq:pdot}) by $-\sqrt{\frac{1}{2m}}$ and add them to find 
\begin{align} 
\label{eq:udot1} 
\dot{iu_{j}} & = i\sqrt{\frac{g}{2mr}}p_{j}+\frac{g}{r}\sqrt{\frac{m}{2}}\left[1+\alpha\left(j-\frac{N}{2}\right)\right]y_{j} \\ \nonumber &+ \sqrt{\frac{1}{2m}}\left[\kappa_{j-1,j}\left(y_{j}-y_{j-1}\right)-\kappa_{j,j+1}r\left(y_{j+1}-y_{j}\right)\right]. 
\end{align}
Next, we define the set of new parameters, $\omega_{0}$, $Ea$ and $t_{j,j+1}$ all with dimensions of one over time, as given in the main text. Substituting $y_{j}=\sqrt{\frac{r}{2mg}}\left(u_{j}+u_{j}^{*}\right)$ into the r.h.s of Eq.~(\ref{eq:udot1}) and using the definitions of $\omega_{0}$, $Ea$ and of $t_{j,j+1}$ from Eq.~(\ref{eq:Mapping-ssh}), we find that
\begin{align}
i\dot{u}_{j} & = \omega_{o}u_{j}+\frac{Ea}{2}\left(j-\frac{N}{2}\right)u_{j}-t_{j-1,j}u_{j-1}+t_{j,j+1}u_{j+1} \\ \nonumber & + A(t_{j-1,j}u_{j-1}^{*},t_{j,j+1}u_{j+1}^{*},aEu_{j}^{*}).\label{eq:udot3}
\end{align}
Note that the r.h.s. contains both $u$(t) but also $u^{*}(t)$, where $A(a,b,c)$ is some linear combination of its arguments. Hence the prefactors of $u_{j-1}^{*},u_{j}^{*}$ and $u_{j+1}^{*}$ are $t$'s and $Ea$ which are all assumed to be much smaller than $\omega_{0}$. Next, we make a transition to a rotating frame, by introducing $u_{j}=\psi_{j}(\tau)e^{-i\omega_{0}\tau}$, and $u_{j}^{*}=\psi_{j}^{*}(\tau)e^{i\omega_{0}\tau}$. This transformation will allow writing an equation for the slow varying variable $\psi_{j}(t)$. Inserting into Eq.~(\ref{eq:udot3}) and multiplying by $e^{i\omega_{0}\tau}$, we find
\begin{align}
i\dot{\psi}_{j} & = \frac{Ea}{2}\left(j-\frac{N}{2}\right)\psi_{j}-t_{j-1,j}\psi_{j-1}-t_{j,j+1}\psi_{j+1} \\ \nonumber & + A(t_{j-1,j}\psi_{j-1}^{*},t_{j,j+1}\psi_{j+1}^{*},aE\psi_{j}^{*})e^{i2\omega\tau}.
\end{align}
In this equation, the terms containing $\psi^{*}$ all have a factor $e^{2i\omega_{0}\tau}.$ Averaging over one period $\tau-\frac{\pi}{\omega_{0}}<\tau'<\tau+\frac{\pi}{\omega_{0}}$, all terms proportional to $\psi_{j}^{*}$ are averaged and can be neglected in the limit of $\omega_{0}\rightarrow\infty$. Note that this statement is exact. However, for any realistic system which has finite $\omega_0$, the mapping is an approximation  whose range of validity should be quantified. Under the above approximations, we finally can write the equation of motion for $\psi$ as Eq.~(\ref{eq:SSH}).

\subsection*{Design of the initial condition}

The initial wave pattern was designed to meet the following requirements: 
\begin{itemize}
\item It should keep the pendula's oscillations in the linear regime.
\item It should have a simple realization (most pendula start at rest, and translations of about 10 pendula). 
\item It should be wide enough in k space (narrow enough in real space) such that the Bloch oscillations can be easily observed in real space.
\item It should be narrow enough in k space to avoid significant dissipation during the Bloch oscillations. 
\item Most of the energy should start in the lower band such that the amount of energy in the upper band will be negligible.
\end{itemize}

After checking several options using the simulations, we chose a shape of a real Gaussian in k space in the lower band corresponding to Experiment 1, of width $\sigma_k=0.63$. The Gaussian in the lower band was sampled according to the discrete values of $k$ in our finite-sized experimental system, and was then transformed back to its dimer components $(u_{k,a}, u_{k,b})$ by the inverse transformation of Eq.~(\ref{eq:u_abk}). We then took a Fourier transform of both components, and finally we took the real value of the resulting function in the pendula space (i.e. pendula at rest), and kept only those values that are bigger than $10\%$ of the maximum value of the pattern (which was chosen to be $10\ cm$ maximal translation of a pendulum). The result was an initial translation of 10 pendula which is given in Table \ref{tab:my_label}, rounded to $1\ mm$. Using the rounded values in this table, we designed the board shown in Fig.~\ref{fig:setup}b. With this initial condition, the initial relative energy in the upper band was demonstrated, both theoretically and experimentally, to be less than $1\%$ in all three experiments. The pendula that were chosen to be translated were pendula number 33-42 in Experiments 1 and 2, and pendula number 37-46 in Experiment 3. These choices were made in order to give the wave enough space to move during the Bloch oscillation without reaching the edge of the system.

\begin{table}
    \centering
\begin{tabular}{|c | c |} 
 \hline
 Pendulum number & Amplitude (cm) \\ [0.5ex] 
 \hline
 1 & 3.4  \\ 
 \hline
 2 & 6.7 \\
 \hline
 3 & 7.7 \\
 \hline
 4 & 10.0 \\
 \hline
 5 & 9.6 \\ [1ex] 
 \hline
 6 & 8.3 \\ [1ex] 
 \hline
 7 & 6.7 \\ [1ex] 
 \hline
 8 & 3.9 \\ [1ex] 
  \hline
 9 & 2.6 \\ [1ex] 
  \hline
 10 & 1.0 \\ [1ex] 
 \hline
\end{tabular}
    \caption{The initial translations of the 10 pendula.}
    \label{tab:my_label}
\end{table}

\subsection*{Relating the couplings $\kappa, \kappa'$ to the experimental design}

\begin{figure}
\centering
\includegraphics[scale=0.4]{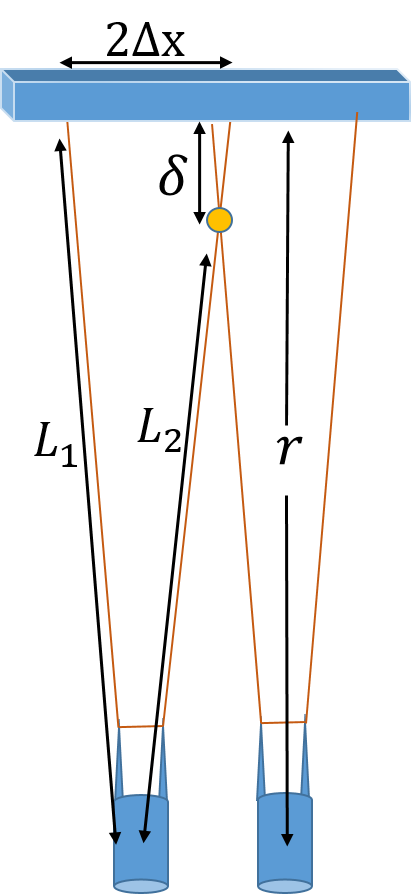}
\caption{The two coupled pendula model used for the estimation of the coupling $\kappa$. }
\label{fig:two_coupled_pendula}
\end{figure}

We analytically estimated $\kappa$ and its dependence on the height of the knots that couple the pendula's strings in the following way. Consider the geometry depicted in Fig.~\ref{fig:two_coupled_pendula}, with only two coupled pendula with equal lengths $r$ and a knot connecting two of their strings at a height $\delta$ below the beam holding the pendula. We calculate $\kappa$  for this model, assuming that the mild change in pendula length in the experiment does not significantly modify the estimation. We compare the dynamic of this two-pendula physical model to an effective system of two pendula attached to one another with a spring of strength $\kappa$ (i.e. as in Eq.~(\ref{eq:pdot}) but with only two pendula). Solving the effective model we find two modes with periodic movements, with frequencies
\begin{align}
\omega_1^2&=\frac{g}{r} , \nonumber\\
\omega_2^2&=\frac{g}{r}+\frac{2\kappa}{m}.\label{eq:omega_2}
\end{align}
The two modes in the physical system (two pendula with a knot attaching two of their wires) are identified as follows: In the first mode, the two pendula move in phase. The knot is meaningless for this mode. The effective length associated with this movement is r, with frequency $\omega_1^2=g/r$. In the second mode, the two pendula move out of phase, at an equal amplitude in opposite directions. In this mode, from symmetry considerations, the joint knot is fixed and doesn't move at all. This mode has a different frequency $\omega_2$ that we solve for by looking at the constraints on the three-dimensional movement $(x,y,z)$ of one of the pendula from its fixed point (say, the left pendula). Two constraints are introduced by its two strings, one string (with length $L_1$) is attached to the fixed beam and the other string (with length $L_2$) is attached to the fixed knot:
\begin{align}
\left(r-z\right)^2+\left(\Delta x+x\right)^2+y^2&=L_1^2 \nonumber \\
\left(r-z-\delta\right)^2+\left(\Delta x+x\right)^2+y^2&=L_2^2   
\end{align}
where $L_1=r^2+\Delta x^2$, $L_2=\left(r-\delta\right)^2+\Delta x^2$, and $\Delta x$ is the distance in the x direction between the attachment points of the two wires, see Fig.~\ref{fig:two_coupled_pendula}. We subtract these two equations and after some algebra we find
\[
z=\frac{2\Delta x}{\delta }x.
\]
As $z$ and $x$ are much smaller than $y$, and as $r$ is significantly longer than any other scale, inserting this relation to any of the above constraints and keeping only linear terms in $z$ leads to
\[
z=\frac{y^2}{2\left(r-\frac{\delta}{2}\right)} 
\]

We should compare this relation to a linear regime of a regular pendulum with an effective length $r_{\rm eff}$, for which $z=\frac{y^2}{2r_{\rm eff}}$ and which leads to a frequency $\omega_2^2=\frac{g}{r_{\rm eff}}$. Hence we find, to first order in $\frac{\delta}{r}$,
\[
\omega_2^2=\frac{g}{r-\frac{\delta}{2}}\approx \frac{g}{r} + \frac{g\delta}{2r^2}
\]
Finally, using Eq.~(\ref{eq:omega_2}) we make the identification
\begin{equation}
\frac{\kappa}{m}=\frac{g\delta}{4r^2}\label{eq:kappa_delta}
\end{equation}

The above estimation was used for the design and for the calibration of the experiment. After the measurements, the couplings $\kappa,\kappa'$ were reexamined using the amplitude of the Bloch oscillations (see further below for the estimation of the amplitude) and the shape of the curve in Fig. \ref{fig:(left)Geometrical phase evollutions} and was found to be in a good agreement with the analytical estimation -- see the section below "Details of the quantitative validation of the mapping" for the details.

\subsection*{Details of the quantitative validation of the mapping }

\subsubsection*{Estimation of the amplitude of Bloch oscillations}\label{sec:estimate_BO_amp}

The initial condition is a Gaussian peaked at some unit cell $\ell_0$. During the Bloch oscillations, the peak of the energy travels to lower unit cells $\ell(t)$ and comes back (Fig.~\ref{fig:(upper-left)-tipycal}). To predict the wave-packet location, we need to keep track of the wavepacket's group velocity. Note that the group velocity changes during the Bloch osculations, according to the dispersion in the two bands, which is given by 
\begin{equation}
\pm\omega_{k}=\pm\left|t+t'e^{-ik}\right|=\pm\left(t^{2}+t'{}^{2}+2tt'\cos k\right)^{1/2}\label{eq:band w_k}
\end{equation}
At half the period of the Bloch oscillation, the wave-packet's center of mass travels to the maximal point of the real space trajectory which is at a distance of 
\begin{align}
\Delta \ell &=\Delta j/2=\left|\int_{0}^{\pi/Ea}\frac{d\omega_{k}}{dk}d\tau\right| \nonumber\\ &=-\int_{0}^{\pi}\frac{d}{dk}\left(t^{2}+t'^{2}+2tt'\cos k\right)^{1/2}\frac{d\tau}{dk}dk\nonumber\\
 & =\frac{2\min(t,t')}{Ea} =\frac{r}{mg\alpha}\min(\kappa,\kappa')
\end{align}
unit cells. As an example, taking the parameters used in Experiments 1 and 2, we obtain an amplitude of $\Delta j\approx6$ pendula, in agreement with the experimentally measured amplitude of the oscillations seen in Fig.~(\ref{fig:(upper-left)-tipycal})c. Taking the parameters in Experiment 3 one obtains an amplitude of $\Delta j\approx15$ pendula.  Note, however, that in Experiment 3, as shown in Fig.~\ref{fig:(upper-left)-tipycal}d, Bloch oscillations were accompanied by a diabetic LZ transition, after which the partial wave in the upper band continued to flow in the same direction. i.e. towards lower $j$. Therefore, the whole journey of the energy from time $\tau=0$ in the lower band to time $\tau=T_B$ in the upper band  required more than $N=2\Delta j\approx30$ pendula (to account also for the width of the wave function and its dispersion in the evolution). Using this calculation, below we make an estimation that in order to fully observe distinguishable Bloch oscillations and LZ transition, $N>40$ is needed, and the Bloch period $T_B$ should be at least as large as 180 times the basic period $2\pi/\omega_0$ of each pendulum.

\subsubsection*{Quantifying the Landau-Zener transition }

In order to quantify the tunneling exponents and timescales associated with the LZ transition, we first need to map the wave dynamics in the SSH problem onto the LZ problem. Expanding Eq.~(\ref{eq:band w_k}) around the point $k=\pi$ in the Brillouin zone where the gap is minimal and the transition occurs, we write $k=\pi+\delta k$ and find
\begin{align*}
\omega_{k} & \approx\pm\left[t^{2}+t'^{2}-2tt'\left(1-\frac{1}{2}\delta k^{2}\right)\right]^{1/2} \nonumber \\ &= \pm\left[(t-t')^{2}+tt'\delta k^{2}\right]^{1/2}.
\end{align*}
As the wavepecket's peak in $k$ space evolves in time at a constant rate and arrives at $k=\pi$ at time $\tau_{0}=\pi/Ea$, we can express the dependence of $\delta k$ on time as $\delta k(\tau)=Ea\left(\tau-\tau_{0}\right)$, so that
\[
\omega_{k}=\pm\left[(t-t')^{2}+tt'\left(Ea\right)^{2}\left(\tau-\tau_{0}\right)^{2}\right]^{1/2}.
\]
We compare this to the two-level Hamiltonian in the standard LZ problem 
$$H=\left(\begin{array}{cc}
\frac{1}{2}d\left(\tau-\tau_{0}\right) & V\\
V & -\frac{1}{2}d\left(\tau-\tau_{0}\right)
\end{array}\right),$$ 
for which the eigenvalues are given by $$\omega_{\rm{LZ}}=\pm\left[\frac{1}{4}d^{2}\left(\tau-\tau_{0}\right)^2+V^{2}\right]^{1/2}.$$ Equating $\omega_{k}$ and $\omega_{\rm{LZ}}$ yields the identification $V=\left|t-t'\right|$, $d=2\sqrt{tt'}Ea$. Using these expressions in the LZ diabatic probability formula, and then converting the SSH parameters into the pendula parameters using Eq.~(\ref{eq:Mapping-ssh}), we find that the diabatic probability can be written as
\begin{align}
P_{D} & = \exp\left(-\frac{2\pi V^{2}}{d}\right) \nonumber \\ &= \exp\left[-\frac{\pi\left(t-t\right)^{2}}{\sqrt{tt'}Ea}\right] \nonumber = \exp\left[-\frac{\pi r\left(\kappa-\kappa'\right)^{2}}{2mg\alpha\sqrt{\kappa\kappa'}}\right] , \label{eq:LZ-pendulum2}
\end{align}
with the r.h.s as in Eq.~(\ref{eq:LZ-pendulum}).

\subsection*{General requirements on the number of pendula and on energy dissipation}

From the previous section a question naturally arises: how large should the system be, and how many pendula oscillations should survive the friction in order to observe the effects studied in this work? Indeed, if one is interested in exploring all effects (both diabatic and adiabatic LZ evolutions, and full Bloch oscillations) while keeping the mapping exact (e.g. keeping $\epsilon<0.2$), then there are general constraints on the system parameters that must be taken into account. 

We first consider the number of pendula needed. Note that if one desires easily distinguishable Bloch oscillations, their amplitude $\Delta j$ should be at least of the order of $\Delta l\gtrsim5$ unit cells, or $\Delta j\gtrsim10$ pendula. In order to observe also the evolution in the upper band after diabatic LZ transitions, one would have to consider twice that number, $2\Delta j$, i.e at least $20$ pendula. In addition to that, one must consider some initial spread of energy across $\sim5$ unit cells, which is an additional $10$ pendula (say in a Gaussian-like shape), in order to have an initial wave packet centered around $k=0$ and of size significantly smaller than the entire Brillouin zone. Lastly, in order to have enough space from the edge of the system to accommodate dispersion in the wave evolution, an additional $5$ pendula on each side should be added. Overall, it seems that $N\gtrsim20+10+10=40$ pendula is a well-suited value for a system in order to observe all effects considered in this paper.

In addition to the above arguments, there are important general constraints and relations between the number of pendula, the steepness of the gradient of the pendula string lengths and the time needed for the energy not to be dissipated from the system. In order to see this, we first note that the gradient of the pendula's length, $\alpha$, has an upper limit given by $N<\frac{2}{\alpha}$, as the length of all pendula must be positive. It is useful to investigate the system using the following dimensionless parameters,
\begin{align*}
\tilde{\tau} =Ea\tau,\quad
\tilde{t}_{j,j+1}=\frac{t_{j,j+1}}{\left(t+t'\right)},\quad
\gamma =\frac{Ea}{\left(t+t'\right)}=\frac{\alpha}{\epsilon^{2}}.
\end{align*}
Using these parameters, with $\tilde{\psi}_{j}(\tilde{\tau})=\psi_{j}(\tau)$,
Eq.~(\ref{eq:SSH}) becomes
\begin{align*}
i \dot{\tilde{\psi}}_{j} & =\frac{\gamma}{2}\left(j-\frac{N}{2}\right)\tilde{\psi}_{j}-\tilde{t}_{j-1,i}\tilde{\psi}_{j-1}-\tilde{t}_{j,i+1}\tilde{\psi}_{j+1},
\end{align*}
with $\tilde{t}_{j,i+1}=\tilde{t}$ ($\tilde{t'}$) for $j$ odd (even) and $\tilde{t}+\tilde{t}'=1$. In this parameterization, one finds that the Bloch oscillations amplitude is given by 
\begin{equation}
\Delta l=\frac{2\epsilon^{2}}{\alpha}\min(\tilde{t},\tilde{t}'),\label{eq:delta_j_2}
\end{equation}
and the diabatic LZ transition probability is given by
\begin{equation}
P_{D}=\exp\left[-\frac{\pi\epsilon^{2}\left(\tilde{t}-\tilde{t}'\right)^{2}}{\alpha\sqrt{\tilde{t}\tilde{t}'}}\right] . \label{eq:P_D2}
\end{equation}
From Eq.~(\ref{eq:delta_j_2}) we must have $N>2\Delta l=\frac{4\epsilon^{2}}{\alpha}\min(\tilde{t},\tilde{t}')$
in order to contain the Bloch oscillations and the diabatic LZ transition within the system size.

Now consider the time needed for the pendula's energy not to dissipate, in order to observe the Bloch oscillations: In order to observe at least one full Bloch period, it means that the Q factor of each pendulum should be of the order of $Q\gtrsim\sqrt{\frac{g}{r}}T_{B}$, where $T_{B}$=$2\pi/Ea$ is the Bloch period. Through the mapping $Ea=\sqrt{\frac{g}{r}}\alpha$, we find $Q\gtrsim2\pi\frac{1}{\alpha}$. How low should $\alpha$ be? As $\tilde{t}+\tilde{t}'=1$, we have $\min(\tilde{t},\tilde{t}')<0.5$, and we immediately find from Eq.~\ref{eq:delta_j_2}) that $\alpha<\frac{\epsilon^{2}}{\Delta l}$. As stated above, if one desires easily distinguishable Bloch oscillations (say $\Delta l\gtrsim5$ unit cells, or $\Delta j\gtrsim10$ pendula), while keeping that mapping exact - say $\epsilon<0.2$, then $\alpha\lesssim0.008$, which gives $Q>800$. Importantly, note that this condition on $\alpha$ and $Q$ is necessary but not sufficient: if we take the above values, $\Delta l>5$, $\alpha=0.008$, $\epsilon<0.2$, we go back to $\min(\tilde{t},\tilde{t}')=\frac{\alpha\Delta l}{2\epsilon^{2}}>0.5$, which in turn leads to $\tilde{t}=\tilde{t}'=0.5$, which in view of Eq.~(\ref{eq:P_D2}) completely closes the gap and leads to $P_{D}=1$. Hence, if one wishes to observe an adiabatic Bloch oscillation in the lower band, $\alpha$ should be lower than $0.008$ and $Q$ bigger than $800$, however not by much: due to the exponential form of Eq.~(\ref{eq:P_D2}), $P_{D}$ decreases very rapidly. For example - if we reduce both $\tilde{t}$ and $\alpha$ by $30\%$, i.e. for $\tilde{t}=0.35$ and $\alpha=0.0056$, Eq.~(\ref{eq:delta_j_2}) keeps $\Delta l=5$ unchanged, while from Eq.~(\ref{eq:P_D2}) we find $P_{D}=0.015$, which is probably satisfactory for any purpose. 

In conclusion, the system requirements are a number of pendula $N\gtrsim40$, whose Q-factor $Q\gtrsim1100$, and which are coupled weakly to each other, such that $\epsilon<0.2$. In our experiment we achieved $Q>2000$, and so we kept $\alpha$ as low as $\alpha=0.0024$ for easier observation of the Bloch oscillation in real space.

\subsection*{Sensitivity to disorder and imperfections in the lattice}

\begin{figure}[b]
\begin{centering}  
\begin{tabular}{cc}
\includegraphics[width=0.5\columnwidth]{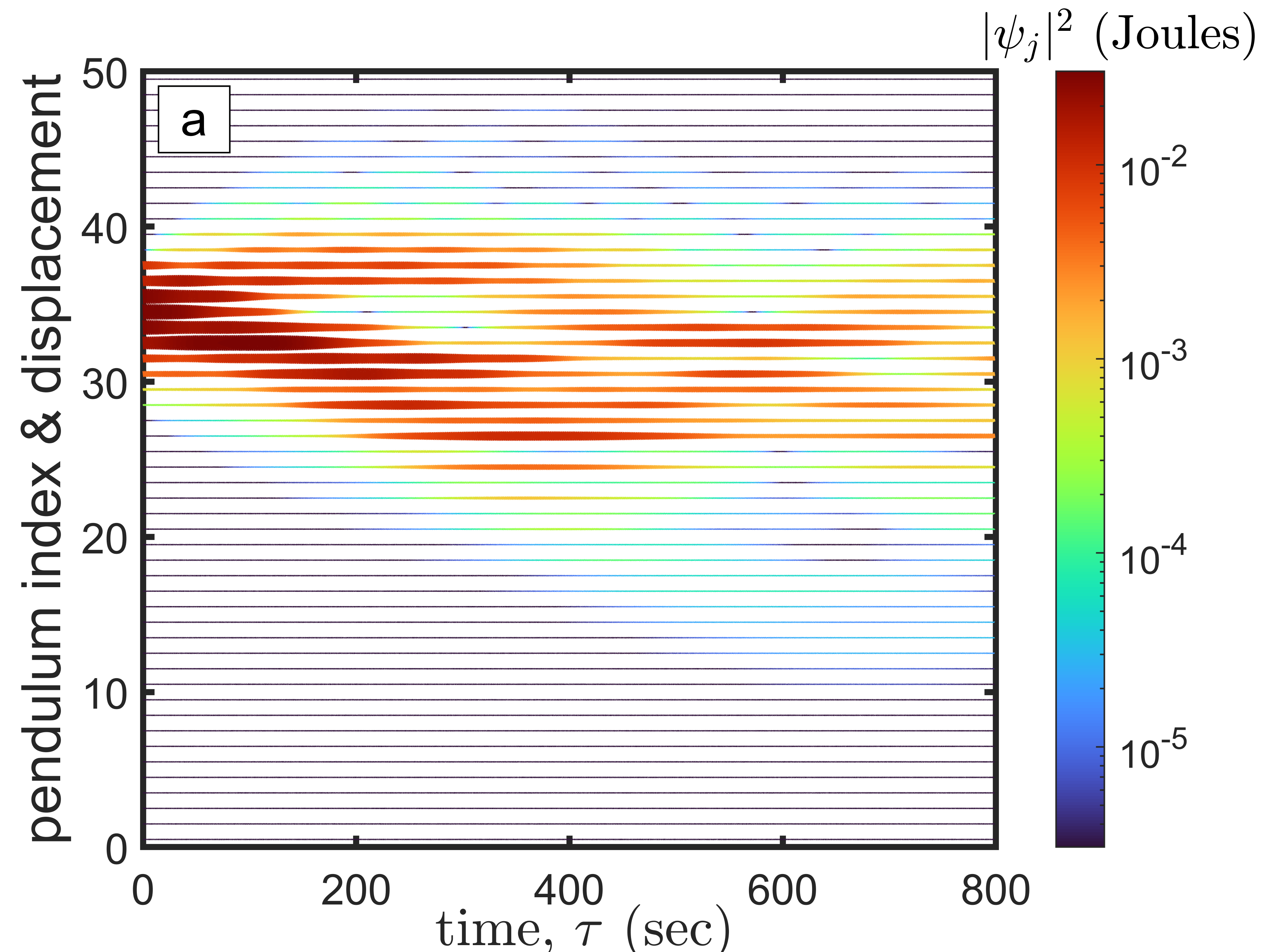} & \includegraphics[width=0.5\columnwidth]{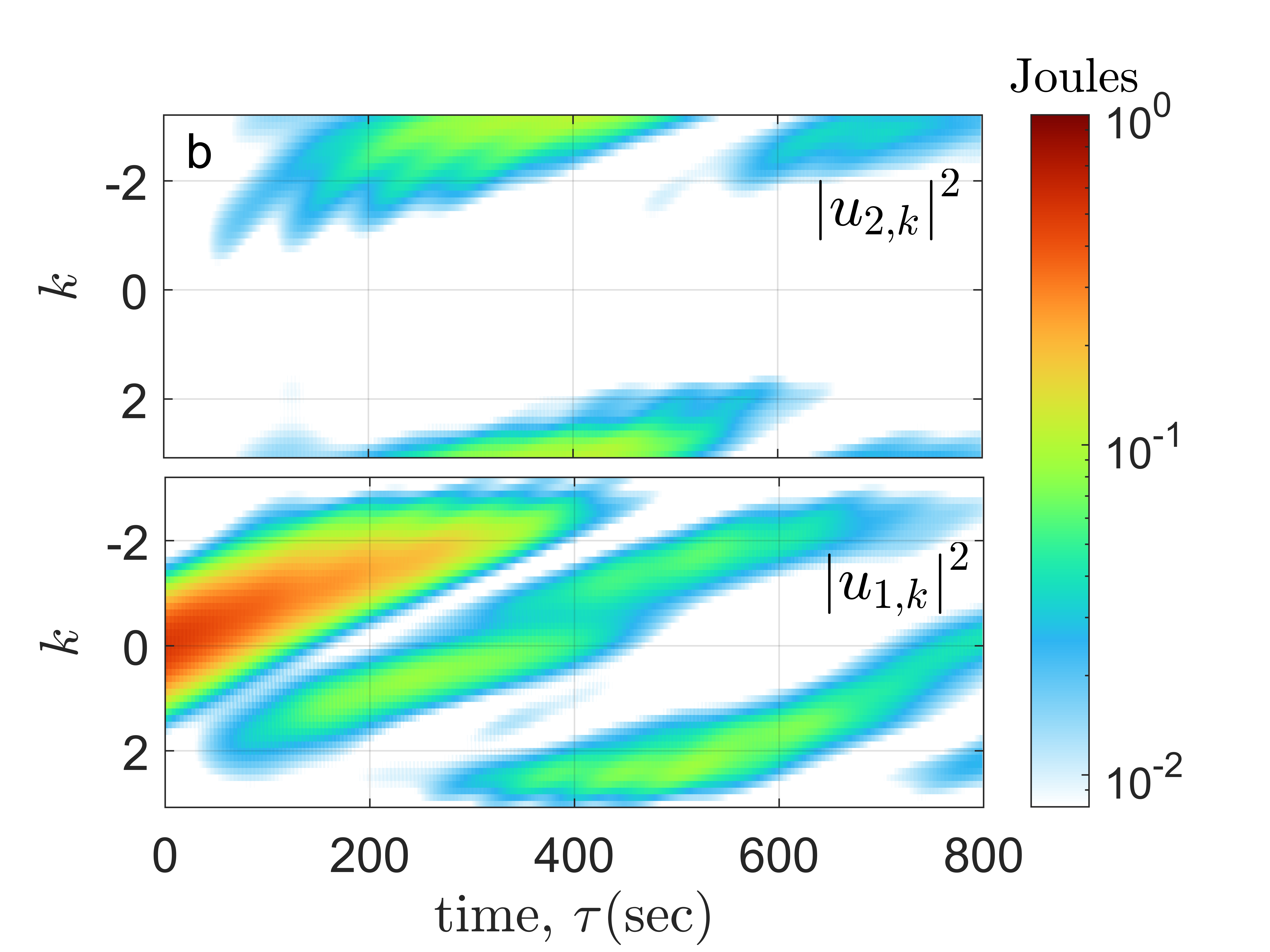}\tabularnewline
\end{tabular}
\par\end{centering}
\caption{The sensitivity of Bloch oscillations to noise in each of the pendula lengths, depicted in real space (a) and in its Fourier transform (b). The parameters of the simulation were as in Experiment 1, but with noise in the lengths, $r_{j}'=r_{j}(1+\eta e_j)$ where $r_{j}$ are the nominal values according to Eq.~(\ref{eq:gradient}) and $\eta=0.0096$. $e_{j}$ is taken from a uniform distribution in the range $[-0.5,0.5]$, and $e_{j}$ and $e_{j'}$ are uncorrelated for $j\protect\neq j'$. \label{fig:The-sensitivity-of_noise}}
\end{figure}

As in the quantum lattice case, also here the Bloch oscillations were found to be highly sensitive to disorder in the lattice, which was considered a risk in measuring the topological phase of the bands. We checked the sensitivity of the Bloch oscillations by running simulations with small random disorder - both in the lengths of the pendula, i.e taking $r_{j}'=r_{j}(1+\eta e_j)$ thus changing the on-site frequency, and in the coupling, namely with $\kappa_{j,j+1}'=\kappa_{j,j+1}(1+\eta e_j)$. In both cases, $\eta$ is the strength of the disorder, and $-0.5 < e_j < 0.5$ are uncorrelated random variables drawn from a uniform distribution.

We first consider the latter case of randomness in the couplings and its effect.  When inserting $\eta=0.1$ relative randomness to each of the couplings $\kappa_{j,j+1}$ (which mimics a realistic calibration error of $2-5$ mm in $\delta_j$ in the experiments) the Bloch oscillations were insensitive to this disorder. However, the phase difference between two successive simulations with switched couplings (as is shown for the experiments in Fig.~\ref{fig:(left)Geometrical phase evollutions}b) showed some sensitivity; the phase evolution deviated from its theoretical value in a random manner which resulted in slight deviation from $\pi$ at the end of the Bloch period. Many realizations of the noise result in a spreading of about $\delta\phi\left(k=0,\tau=T_B\right)=0.18$ (one standard deviation) around the mean value of $\pi$.

\begin{figure}[h]
\begin{centering}
\includegraphics[width=0.8\columnwidth]{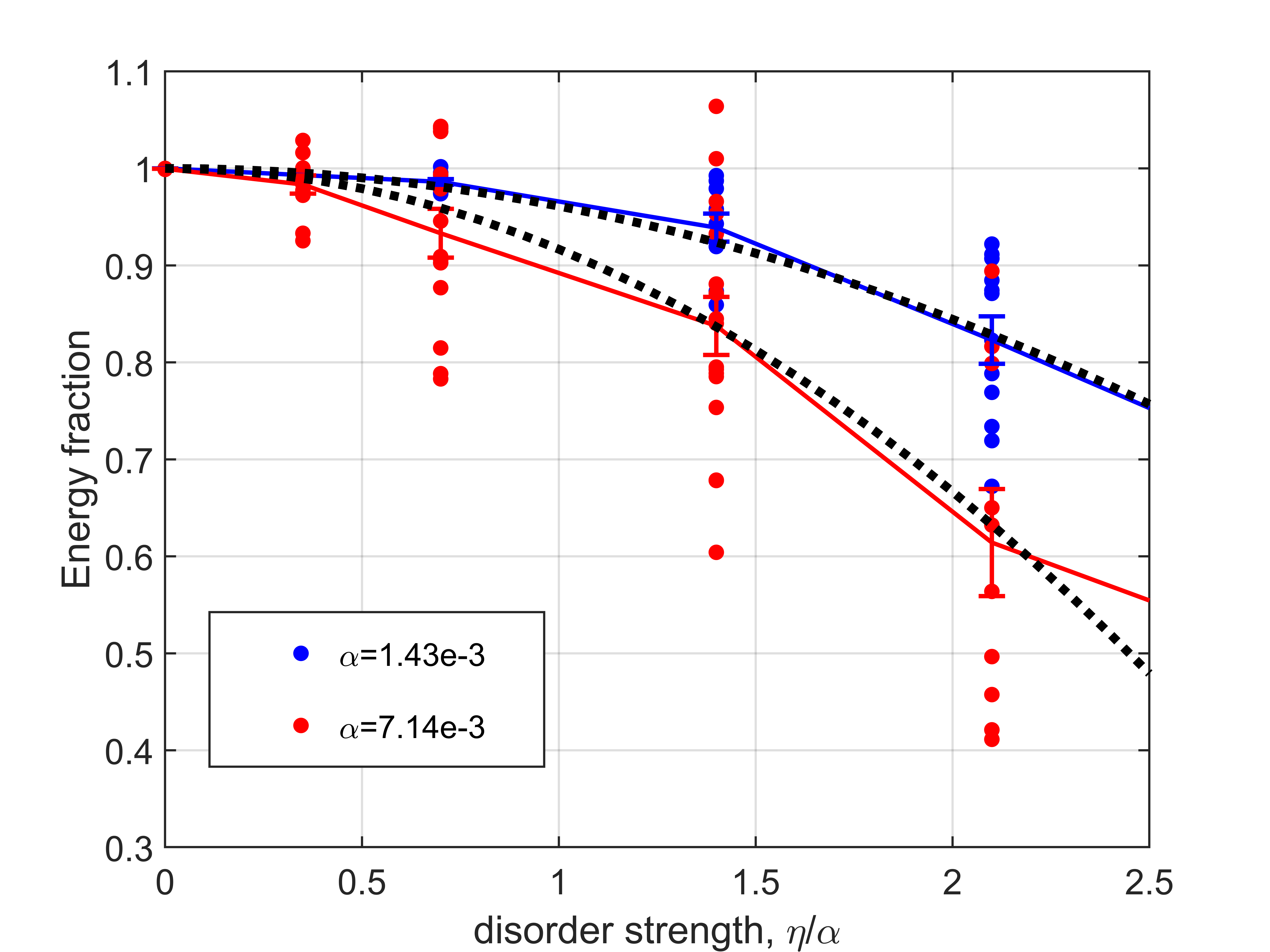}
\par\end{centering}
\caption{Quantification of the effect of the noise on Bloch oscillations survival using many noise realizations. Each point is the result of a simulation with one noise realization, and shows the amount of energy that returns to $\left|k\right|<0.5$ in the lower band after exactly one Bloch period, relative to the energy in the same range of $k$ at $t=0$. Simulations were performed for different values of noise strength and two different values of $\alpha$, which controls the Bloch period, marked by the color of each dot. Solid lines connect the average values for each noise strength, and error bars are the statistical error. Black dashed lines are the function $P=1-\frac{Ea}{2\left(t+t'\right)}\left(\frac{\eta}{\alpha}\right)^{2}$ for the two parameter sets of the simulations. \label{fig:Quantification-of-the noise}}
\end{figure}

The Bloch oscillations were very sensitive to the modifications in the on-site frequency. Adding small noise to $r_{j}$ (of the order of $\alpha r_{j}$) can destroy the Bloch oscillation completely. In Fig.~\ref{fig:The-sensitivity-of_noise} we show the effect of randomizing the lengths $r_{j}$ with $\eta=0.0096$, using a simulation that mimicked Experiment~1. An imperfect preparation of the pendula strings lengths of less than one percent results in a significant scrambling of $k$ within a single period of the Bloch oscillation, which leads to vanishing of the Bloch oscillations. In Fig.~\ref{fig:Quantification-of-the noise} we show the effect of many realizations of the noise on the scrambling of $k$ during one period, as a function of noise strength. It appears that the average of those realizations follow the formula $P=1-\frac{Ea}{2\left(t+t'\right)}\left(\frac{\eta}{\alpha}\right)^{2}=1-\frac{\epsilon^2\eta^2}{\alpha}$, but this result lacks a theoretical explanation.
 
\subsection*{Derivation of the phase evolution in Fourier space}

We wish to predict theoretically the phase evolution of the wave function in the lower band in the case of a completely adiabatic evolution, given that the Bloch oscillations cause the wave function to explore the entire Brillouin zone during one period. To this end, we start with the SSH model with an electric field on a lattice, Eq.~(\ref{eq:SSH}):
\[
i\dot{\psi}_{j}=\frac{Ea}{2}\left(j-\frac{N}{2}\right)\psi_{j}-t_{j,j+1}\psi_{j+1}-t_{j-1,j}\psi_{j-1} .
\]
Note that $N$ is the total number of pendula. We shall assume that the wave does not reach the edges of the system, so we have $\psi_{1}=\psi_{N}=0$ and we can safely fix $\psi_{0}=\psi_{N+1}=0$. In order to identify the two bands, we switch to even-odd language (a,b) with $l=\left\lfloor \left(j+1\right)/2\right\rfloor$ counting the unit cells and with $t\equiv t_{j,j+1}$ for odd $j$ and $t'\equiv t_{j,j+1}$ for even $j$. With this parameterization, we get the two equations
\begin{align}
i\dot{\psi}_{a,l} & =-\frac{Ea}{2}\psi_{a,l}+Ea\left(l-\frac{N}{4}\right)\psi_{a,l}-t\psi_{b,l}-t'\psi_{b,l-1},\nonumber\\
i\dot{\psi}_{b,l} & =+Ea\left(l-\frac{N}{4}\right)\psi_{b,l}-t'\psi_{a,l+1}-t\psi_{a,l}.\label{eq:ssh_lab_space}
\end{align}
We now perform a discrete Fourier transform, by multiplying all terms by $e^{-ikl}$ and summing over $l$. As $le^{-ikl}=i\frac{d}{dk}e^{-ikl}$, we find the following equations, written in matrix form:
\begin{align}
i\left(\begin{array}{c}
\dot{\psi}_{a,k}\\
\dot{\psi}_{b,k}
\end{array}\right) &= Ea\left(i\frac{d}{dk}-\frac{N}{4}\right)\left(\begin{array}{c}
\psi_{a.k}\\
\psi_{b.k}
\end{array}\right) \nonumber \\ &-\left(\begin{array}{cc}
Ea/2 & t+t'e^{-ik}\\
t+t'e^{ik} & 0
\end{array}\right)\left(\begin{array}{c}
\psi_{a.k}\\
\psi_{b,k}
\end{array}\right)\label{eq:almost SSH}
\end{align}

At this point, a few approximations can be made to simplify the derivation below. First, we neglect the difference of $Ea/2$ between the diagonal entries in the matrix in Eq.~(\ref{eq:almost SSH}). It originates from the fact that the potential term $\frac{Ea}{2}(j-\frac{N}{2})$ has different values at the $a$ and $b$ sites, so the matrix is not strictly off-diagonal. However, in our simulation we kept $\frac{Ea}{4}\ll t,t'$ such that the diagonal terms of the matrix, $\pm\frac{Ea}{4}$ added a negligible contribution to the phase evolution after one Bloch period, proportional to $\frac{\pi}{2}\cdot\frac{Ea}{\min(t,t')}$, which was less than 0.1 radians. 

We also neglect any constants added to the first term of order $Ea$. A term $Ea/4$ is due from the above-mentioned and neglected $Ea/2$ diagonal term in the matrix above, but it can also originate for example from a different choice of the Fourier transform's phase of exponential, i.e taking $e^{-ik(l-1)}$ instead of $e^{-ikl}$ in the transition from Eq.~(\ref{eq:ssh_lab_space}) to Eq.~(\ref{eq:almost SSH}). Such terms would contribute a globally oscillatory term to the wave function, hence a small linear contribution to the phase evolution, which does not affect any of the results given below. 

Neglecting the diagonal term in the matrix in Eq.~(\ref{eq:almost SSH}), we obtain the equation
\begin{align}
i\left(\begin{array}{c}
\dot{\psi_{a,k}}\\
\dot{\psi_{b,k}}
\end{array}\right) &= Ea\left(i\frac{d}{dk}-\frac{N}{4}\right)\left(\begin{array}{c}
\psi_{a,k}\\
\psi_{b,k}
\end{array}\right) \nonumber \\ &-\left(\begin{array}{cc}
0 & t+t'e^{-ik}\\
t+t'e^{ik} & 0
\end{array}\right)\left(\begin{array}{c}
\psi_{a,k}\\
\psi_{b,k}
\end{array}\right)\label{eq: truely SSH}
\end{align}
that we wish to solve. We first diagonalize the matrix for general $k$ to get the two eigenvalues using the unitary transformation from $(a,b)$ to $(1,2)$,
\begin{equation}
\left(\begin{array}{c}
\psi_{a,k}\\
\psi_{b,k}
\end{array}\right)=\frac{1}{\sqrt{2}}\left(\begin{array}{cc}
1 & 1\\
e^{i\varphi_{k}} & -e^{i\varphi_{k}}
\end{array}\right)\left(\begin{array}{c}
\psi_{1,k}\\
\psi_{2,k}
\end{array}\right)\label{Eq. unitary},
\end{equation}
where $\varphi_{k}$ is the phase of the off-diagonal term, $t+t'e^{ik}\equiv v_{k}e^{i\varphi_{k}}$. Performing this transformation on all terms of the above equation, we have
\begin{align}
i\left(\begin{array}{c}
\dot{\psi}_{1,k}\\
\dot{\psi}_{2,k}
\end{array}\right)_{k}= & Ea\left(i\frac{d}{dk}-\frac{N}{4}\right)\left(\begin{array}{c}
\psi_{1,k}\\
\psi_{2,k}
\end{array}\right) \nonumber \\ &+\left(\begin{array}{cc}
-v_{k} & 0\\
0 & v_{k}
\end{array}\right)\left(\begin{array}{c}
\psi_{1,k}\\
\psi_{2,k}
\end{array}\right)\nonumber \\
 & -Ea\frac{\varphi_{k}'}{2}\left(\begin{array}{cc}
1 & -1\\
-1 & 1
\end{array}\right)\left(\begin{array}{c}
\psi_{1,k}\\
\psi_{2,k}
\end{array}\right)\label{eq: SSH 1,2}.
\end{align}
Note that the last term is the result of the derivative of the unitary transformation in Eq.~(\ref{Eq. unitary}) with respect to $k$. In addition, note that from this result one can understand the significance of the minus sign of the two coupling terms in Eq.~(\ref{eq:SSH}); it causes the lower band's eigenstate at $k=0$ to be $\xi_{1,k=0}=\frac{1}{\sqrt{2}}(1,1)^{T}$ and the state for the upper band is $\xi_{2,k=0}=\frac{1}{\sqrt{2}}(1,-1)^{T}$, and not the other way around. This made the preparation of the initial wave packet in the lower band easier, as all the pendula initially began with the same sign of displacement from their equilibrium positions.

Next, we solve for the case of the fully adiabatic limit, i.e. for a large band gap, $\left|t-t'\right|\gg Ea$. Knowing the adiabatic solution to the LZ problem for this case, we assume a solution confined only to the lower band, namely of the form
\begin{equation}
\left(\begin{array}{c}
\psi_{1,k}(\tau)\\
\psi_{2,k}(\tau)
\end{array}\right)=\left(\begin{array}{c}
f(s)e^{-i\phi(\tau,s)}\\
0
\end{array}\right),\label{eq:anzats}
\end{equation}
where $s=k+Ea\tau$. This form assumes that the positive amplitude of the wave is $f(k)$, taken from the initial condition at $\tau=0$, is constant and does not change in time, except for a shift in $k$ due to the Bloch oscillations. In the case that $f(k)$ is centered at $k_{c}(0)=0$ at $\tau=0,$ the wave packet at later time $\tau$ is centered around $k_{c}(\tau)=-Ea\tau$. 

The global phase evolution $\phi(\tau,s)$ has to be determined from Eq.~(\ref{eq: SSH 1,2}). Changing variables from $\left(\tau,k\right)$ to $\left(\tau,s\right)$, we use the relations between partial derivatives:
\begin{align*}
i\left(\left.\frac{\partial}{\partial\tau}\right|_{k=constant}-Ea\left.\frac{\partial}{\partial k}\right|_{\tau=const}\right) & =i\left.\frac{\partial}{\partial\tau}\right|_{s=const}
\end{align*}

Using this in Eq.~(\ref{eq: SSH 1,2}) and then inserting Eq.~(\ref{eq:anzats}) and therefore neglecting any $\psi_{2}$ terms, we find in the ($s,\tau$) coordinates that
\begin{align*}
\frac{\partial\phi(s,\tau)}{\partial\tau} & =\left(-Ea\frac{N}{4}-v_{k(s,\tau)}-Ea\frac{\varphi'_{k(s,\tau)}}{2}\right),
\end{align*}
with $\varphi'_{k(s,\tau)}\equiv\left.\frac{\partial\varphi_{k}}{\partial k}\right|_{k=s-Ea\tau}$. Next, integrating the above equation over $\tau$ at constant $s$ leads to the solution of the phase evolution in Fourier space:
\[
\phi(s,\tau)=\phi_{0}(s)+\int_{0}^{\tau}\left(-Ea\frac{N}{4}-v_{k(s,\tau')}-Ea\frac{\varphi'_{k(s,\tau')}}{2}\right)d\tau'.
\]
Changing the solution back to the $(k,\tau)$ coordinate system and using $Ea\varphi'_{k(s,\tau')}d\tau'\equiv-\frac{\partial\varphi_{k}}{\partial k}dk$, we find
\begin{align}
\phi(k,\tau) &= \phi_{0}(k+Ea\tau)-\frac{N}{4}Ea\tau \nonumber \\ &-\int_{0}^{\tau}v_{k'=k+Ea(\tau-\tau')}d\tau'+\frac{\left.\varphi_{k}\right|_{k+Ea\tau}^{k}}{2},\label{eq:phi_tau_k}
\end{align}
which is Eq.~(\ref{eq:phase evolution}) in the main text. One can identify the meaning of the four terms in the r.h.s of this equation: 
The last two terms are the dynamic phase and geometrical phase:
\begin{align*}
\phi_{d}(k,\tau) & \equiv-\int_{0}^{\tau}v_{k'=k+Ea(\tau-\tau')}d\tau'\\
\phi_{g}(k,\tau) & \equiv\frac{\left.\varphi_{k}\right|_{k+Ea\tau}^{k}}{2}.
\end{align*}
The second term is the contribution of the potential zero shift from $j=0$ to $j=N/2$, which is $l=N/4$ (given that the phase in the Fourier transform is taken from the $l=0$ unit cell). 
The function $\phi_{0}$ in the first term is determined from the initial shift of the wave's symmetry axis at $\tau=0$. To see this, note that if we shift the initial wave function at $\tau=0$ by an even number of pendula $j_{0}$ about $j=0$ (i.e. by $l_{0}=j_{0}/2$ unit cells), the Fourier wave function gets an oscillating prefactor, such that
\[
\psi_{1,k}(\tau=0)=e^{ikl_{0}}r_{k}.
\]
In the experiment, we aimed for a real $r_k$ (a Guassain in the lower band). In view of Eq.~(\ref{eq:anzats}), this initial condition leads to $\phi_{0}(k+Ea\tau)=(k+Ea\tau)l_{0}.$

\end{document}